\documentclass[aip,graphicx,reprint]{revtex4-1}

\usepackage[dvipsnames]{xcolor}
\usepackage{graphicx}

\draft

\begin{document}


\title{Tunable Giant Rashba-type Spin Splitting in PtSe$_2$/MoSe$_2$ Heterostructure}
\author{Longjun Xiang}
\author{Youqi Ke}
\email{keyq@shanghaitech.edu.cn}
\author{Qingyun Zhang}
\email{zhangqy2@shanghaitech.edu.cn}
\affiliation{School of Physical Science and Technology, ShanghaiTech University, Shanghai, 201210, China}
\affiliation{University of Chinese Academy of Sciences, Beijing 100049, China}
\date{\today}

\begin{abstract}
We report a giant Rashba-type spin splitting in two-dimensional heterostructure PtSe$_2$/MoSe$_2$ with first-principles calculations. We obtain a large value of spin splitting energy 110 meV at the momentum offset $k_0$=0.23 \AA$^{-1}$ around $\mathrm{\Gamma}$ point, arising from the emerging strong interfacial spin-orbital coupling induced by the hybridization between PtSe$_2$ and MoSe$_2$. Moreover, we find that the band dispersion close to valence band  maximum around $\Gamma$ point can be well approximated by the generalized Rashba Hamiltonian $H(k_{||})=-\frac{\hbar^2 k_{||}^2}{2m}+c k_{||}+\alpha_R \vec{\sigma}\cdot(\vec{k}_{||} \times \vec{z})$. It is found that the generalized Rashba constant $\eta_R=c+\alpha_R$ in PtSe$_2$/MoSe$_2$ is as large as 1.3 eV$\cdot\text{\AA}$, and importantly $\eta_R$ can be effectively tuned by biaxial strain and external out-of-plane electrical field, presenting a potential application for the spin field-effect transistor. In addition, with the spin-valley physics at $\mathrm{K}/\mathrm{K}'$ points in monolayer MoSe$_2$, we propose a promising model for spin field-effect transistor with opto-valleytronic spin injection based on PtSe$_2$/MoSe$_2$ heterostructure.
\end{abstract}


\maketitle 

Finding materials with strong spin-orbit coupling (SOC) is at the heart of the research field of spin-orbitronics.\cite{Manchon:2015}  For example, the Datta-Das spin field-effect transistor (SFET) \cite{Datta:1990} requires the channel material with large Rashba-type spin splitting and electrically tunable Rashba constant\cite{Rashba:1984} to effectively modulate the electron current by steering the spin precession.
The original Rashba Hamiltonian for two-dimensional (2D) electron gas is given by $H_R = \pm \frac{\hbar^2 k_{||}^2}{2m} + \alpha_R\vec{\sigma}\cdot(\vec{k}_{||} \times \vec{z})$, in which
the Rashba constant $\alpha_R$ represents the strength of SOC and $\vec{\sigma}$ the vector of Pauli matrices.
In the past decades, considerable theoretical and experimental efforts have been devoted to
finding Rashba SOC at interfaces or surfaces. The Rashba-type spin splitting
was observed in III-V semiconducting
heterostructure InGaAs/InAlAs\cite{Nitta:1997} with tiny Rashba constant
($\alpha_R$=$0.001$ eV$\cdot$\text{\AA}).
The surface states of
heavy metals (\textit{e.g.}, Au,\cite{LaShell:1996} Bi,\cite{Koroteev:2004,Hirahara:2006}
Ir,\cite{Ir:2012} Pb\cite{Hugo:2008})
and topological insulator Bi$_2$Se$_3$\cite{King:2011,Zhu:2011} exhibited
large Rashba-type spin splittings.
In addition, surface alloys, by doping heavy atoms at the surface of metals
(\textit{e.g.}, Bi/Ag(111)\cite{Bi_Ag111:2007}) or semiconductors
(\textit{e.g.}, Bi/Si(111)\cite{Bi_Si111:2009} and Cs/InSb(110)\cite{Bindel:2016}),
provided an effective strategy to generate
giant Rashba-type spin splitting. Typically, these observed
Rashba-type surface states are accompanied with trivial surface states and spin degenerate bulk states,
which hinders their practical applications. Recently, giant Rashba-type spin splitting is
also observed in bulk BiTeI\cite{Ishizaka:2011} and GeTe.\cite{Sante:2013,Liebmann:2016} However, it is unclear
whether these effects can survive when the bulk materials
are downsized to a few atomic
layers.\cite{Ma:2014}  After the discovery of graphene, 2D materials
\begin{figure}[t!]
\centering
\includegraphics[width=0.7\columnwidth]{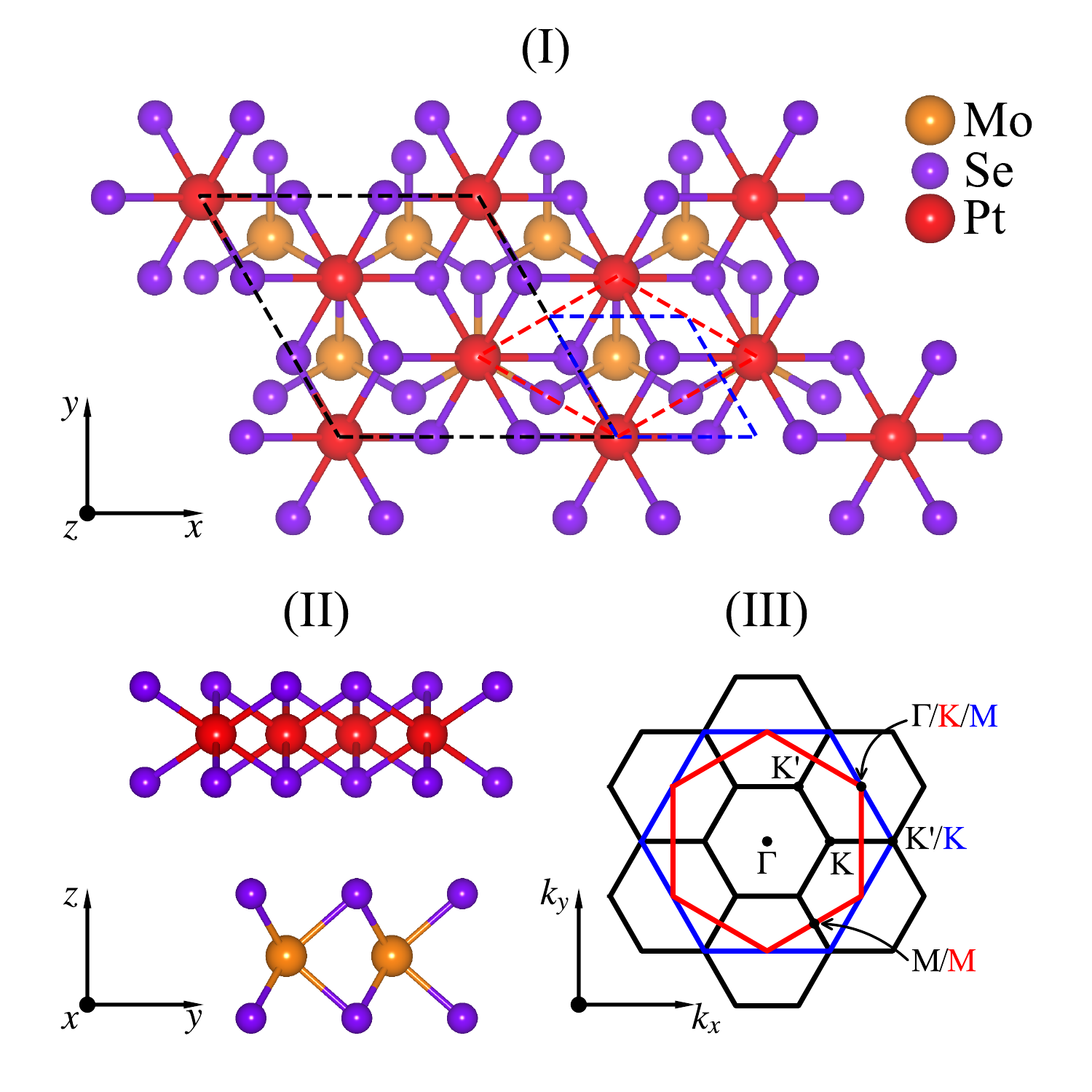}
\caption{\label{struct} (I) The top and (II) side view of the heterostructure PtSe$_2$/MoSe$_2$.
(III) The corresponding
Brillouin zones for the PtSe$_2$/MoSe$_2$ heterostructure and their constituent monolayers.}
\end{figure}
\begin{figure*}[t!]
\centering \includegraphics[width=1.7\columnwidth]{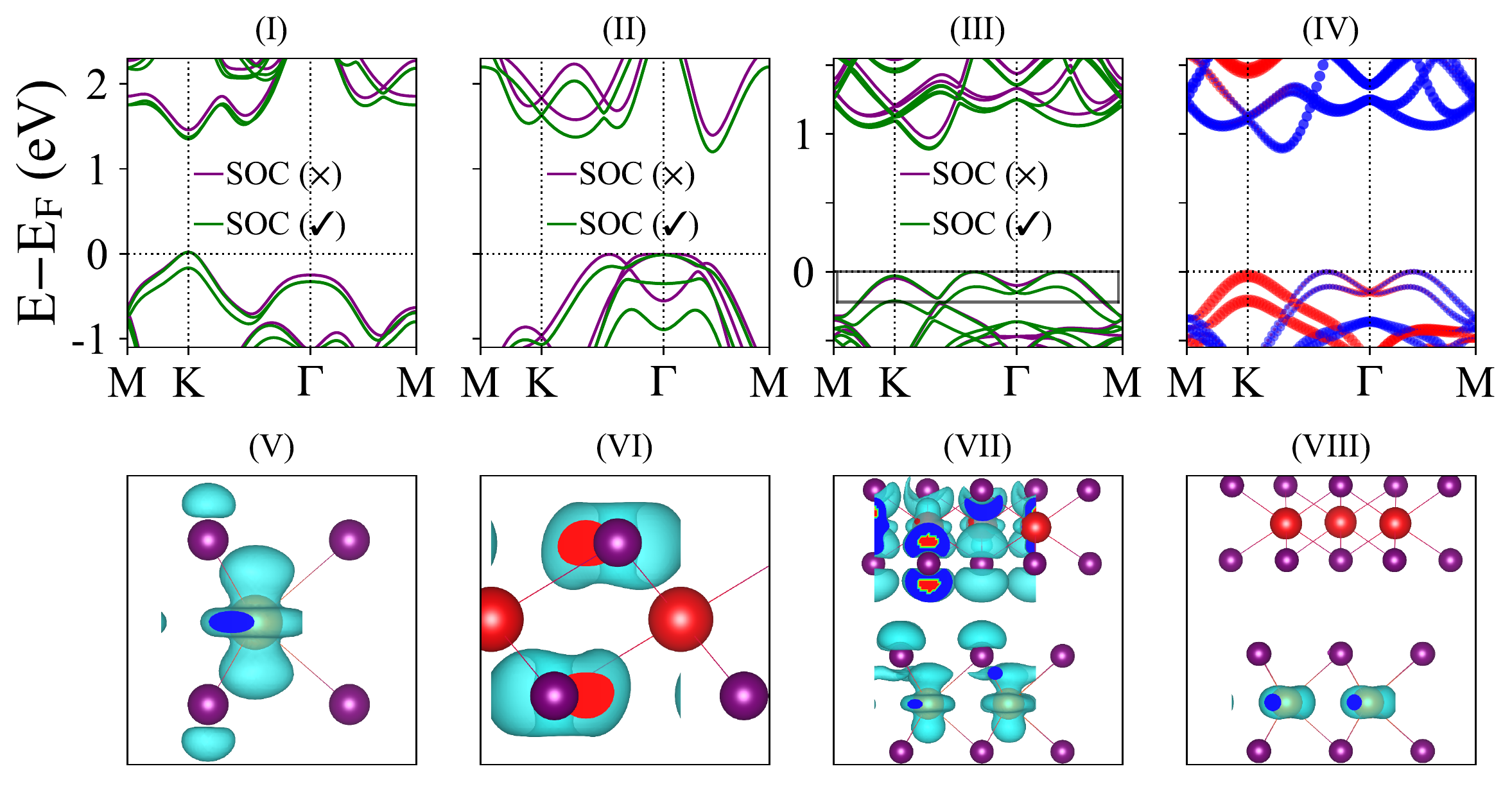}
\caption{\label{proj_band_sum} Band structures of (I) monolayer MoSe$_2$,
(II) monolayer PtSe$_2$,
and (III) hetrostructure PtSe$_2$/MoSe$_2$ without (purple) and with (green) SOC.
(IV) Layer-projected band structure for PtSe$_2$/MoSe$_2$,
in which the red and blue circles represent the respective contributions from MoSe$_2$ and PtSe$_2$.
The charge densities of VBE around $\Gamma$ point for
(V) monolayer MoSe$_2$,
(VI) monolayer PtSe$_2$,
(VII) heterostructure PtSe$_2$/MoSe$_2$,
and (VIII) at $\mathrm{K}$ point for PtSe$_2$/MoSe$_2$.
The isosurface values of 0.005 and 0.0005 electron/\text{\AA}$^3$ are adopted for monolayers
and heterostructure, respectively.}
\end{figure*}
and their van der Waals heterostructures\cite{Ajayan:2016,Novoselov:2016,Kasai:2018} offer
new opportunities to search for materials with Rashba SOC.\cite{Manchon:2015}
For example, the Rashba effect was predicted in the monolayer LaOBiS$_2$ \cite{Liuqihang:2013} 
and BiSb \cite{Singh:2017}. By stacking Bi$_2$Se$_3$ ultrathin film on MoTe$_2$ substrate, Wang \textit{et at.}\cite{Wang:2017}
predicted a wide-range Rashba electron gas. 
The Rashba effect was also predicted in other van der Waals heterostructures.\cite{Singh:2017,Zhang:2018} 
Despite these important progresses, challenges still exist in these predicted systems,
such as the accompanying non-Rashba states, weak tunability of Rashba constant and small Rashba energy $E_R$ 
(for example, $8$ meV in Bi$_2$Se$_3$/MoTe$_2$\cite{Wang:2017}, $5$ meV in BiSb/AIN\cite{Singh:2017}.).
Therefore, further efforts are required to search for appropriate materials with tunable Rashba SOC and large spin splitting.

Very recently, the semiconducting T-phase transition metal dichalcogenide (TMDC) PtSe$_2$ monolayer
has attracted much attention due to its superior transport properties.\cite{Wang:2015,Zhao:2017}
Moreover, Yao \textit{et al.}\cite{Yao:2017} observed the spin-layer locking
phenomena\cite{Zhang:2014,Yuan:2019} induced by local dipole field in this material,
which manifests the significant role
of SOC. However, the spin degeneracy protected by inversion symmetry in PtSe$_2$
impedes its applications to SFET.
To lift the degeneracy the structural inversion asymmetry should be introduced,
for example, by substitutional doping.\cite{PtSe2_doping:2018}
However, we note that building van der Waals heterostructure
is more favorable in experiment compared to the doping method.
On the other hand, monolayer H-phase TMDCs, such as MoSe$_2$, have been exfoliated
successfully and also possess strong SOC.\cite{Chang:2014,Reyes:2016}
Therefore, the heterostructure stacked by T-phase and H-phase TMDCs monolayer, with broken
inversion symmetry, provides a promising platform for studying spin physics.
In this work, we investigate the electronic properties of the heterostructure
PtSe$_2$/MoSe$_2$ through first-principles calculations.
We demonstrate that a generalized Rashba-type electron system can be formed and effectively tuned 
by exerting biaxial strain and external out-of-plane electrical field, 
providing a promising material for SFET application.

All the calculations are performed with density functional theory (DFT) by the
projector-augmented wave (PAW) method implemented in the Vienna Ab initio Simulation
Package (VASP)\cite{Kresse:1999}. The generalized gradient approximation as parameterized
by Perdew, Burke, and Ernzerhof (PBE)\cite{Perdew:1996} is employed. The kinetic energy
cutoff $460$ eV and a $\mathrm{\Gamma}$-centered $6 \times 6 \times 1$ k-mesh in $2$D
Brillouin zone have been used for the convergence of total energy with a criterion of
1$\times$10$^{-6}$ eV per atom. The SOC is not included during the geometry optimization
but is added in the electronic structure calculations. The lattice constants of monolayer
MoSe$_2$ and PtSe$_2$ are 3.286 \text{\AA} and 3.780 \text{\AA}, respectively.
To minimize the artificial internal strain caused by lattice mismatch, the heterostructure
is built from $\sqrt{3}\times \sqrt{3}\times 1$ supercell of PtSe$_2$ and $2 \times 2 \times 1$
supercell of MoSe$_2$, as shown in FIG.\ref{struct} (I). A vacuum space of 20 \text{\AA}
is adopted to avoid the interaction between two periodic
slabs along $[001]$ direction. The DFT-D3\cite{Grimme:2010}
method has been applied to simulate the van der Waals interaction between MoSe$_2$ and PtSe$_2$
(see TABLE S1 in the supplement for more structural information).

In FIG.\ref{proj_band_sum} (I-III), we present the band structures for monolayer MoSe$_2$, PtSe$_2$, and their heterostructure without/with SOC. 
The Fermi level is set to valence band maximum (VBM) in all band structures. As shown in FIG.\ref{proj_band_sum} (I)-(II), the pristine monolayers
MoSe$_2$\cite{Schwingen:2011} and PtSe$_2$\cite{Wang:2015} are direct and indirect semiconductors,
respectively. When SOC is considered, the bands of monolayer MoSe$_2$ (PtSe$_2$)
are split (not split) along $\mathrm{M}$-$\mathrm{K}$-$\mathrm{\Gamma}$ $k$-path
due to the absence (presence) of inversion symmetry.
Because the heterostructure PtSe$_2$/MoSe$_2$
is formed by the interlayer van der Waals interaction which is relatively weaker than intralayer bonding,
therefore, the band structure of PtSe$_2$/MoSe$_2$ should preserve most of the characteristics of
its constituent monolayers. As expected,
the heterostructure PtSe$_2$/MoSe$_2$
retains the semiconducting nature with an indirect band gap $0.89$ eV when SOC is considered,
as shown in FIG.\ref{proj_band_sum} (III).
In addition, the heterostructure also preserves the large spin splitting (181 meV)
at $\mathrm{K}$ point with a slight reduction in comparison with that in monolayer MoSe$_2$ (186 meV as seen in FIG.\ref{proj_band_sum} (I)).

\begin{figure}[t!]
	\centering \includegraphics[width=0.8\columnwidth]{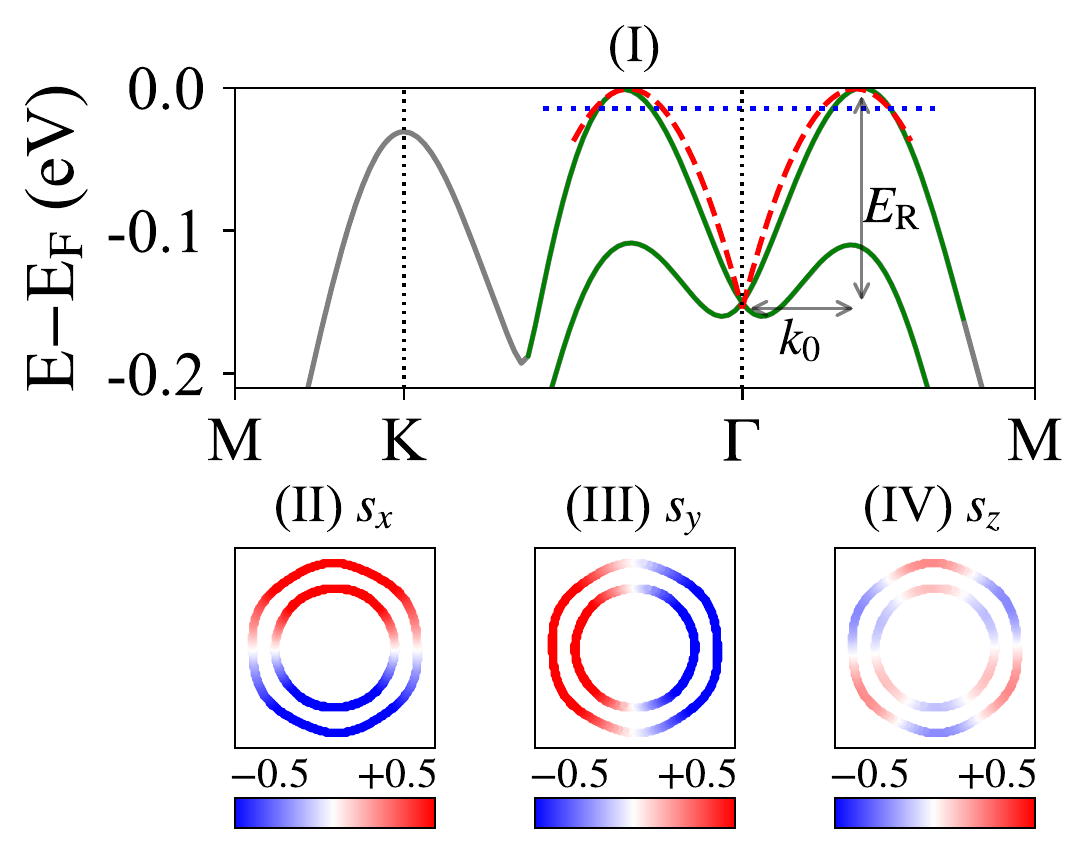}
	\caption{
		\label{resolved_spintxt}
		(I) The two highest valence bands. The upper band between 0 eV to -0.1 eV around $\Gamma$ point can be
		described by a generalized Rashba Hamiltonian (red dashed line). 
		(II)-(IV) Spin components around $\Gamma$ point for a constant energy cut (blue dotted line in (I)).
	}
\end{figure}

Interestingly, an emerging spin splitting around $\mathrm{\Gamma}$ point is observed,
which does not appear in the monolayer MoSe$_2$ and PtSe$_2$.
To unveil its physical origins, we firstly
investigate the layer-projected band structure, as shown in FIG.\ref{proj_band_sum} (IV).
The contribution from the layer MoSe$_2$ and PtSe$_2$ are indicated by
red and blue circles, respectively.
From the projected band structure, we find an important hybridization from two constituent monolayers
at the valence band edge (VBE) around $\mathrm{\Gamma}$ point.
Compared to previous studies of bilayer MoS$_2$ and heterostructure MoSe$_2$/MoS$_2$\cite{Kang:2013,Komsa:2013,Su:2016},
in which the hybridization around $\Gamma$ point is dominated by Mo-$d$ orbitals
(see FIG.S2 in the supplement.),
the observed hybridization around $\Gamma$ point in PtSe$_2$/MoSe$_2$ displays a distinct feature:
it is mainly contributed by Mo-$d$ orbitals in MoSe$_2$ and Se-$p$ orbitals in PtSe$_2$.
Furthermore, to capture more details of
the band splitting around $\mathrm{\Gamma}$ point,
the band decomposed charge densities are plotted, as shown in FIG.\ref{proj_band_sum} (V-VII).
For monolayer PtSe$_2$ and MoSe$_2$, the charge density around $\Gamma$ point has an equal contribution from top and bottom Se layers.
Therefore, the total vertical electrical field acting on the electrons is canceled and no band splitting can be observed. 
However, for PtSe$_2$/MoSe$_2$ heterostructure as shown in FIG.\ref{proj_band_sum} (VII), the charge density around $\mathrm{\Gamma}$ point, 
mainly contributed by $d_{z^2}$ of Mo in MoSe$_2$ and $p_z$ of Se in PtSe$_2$,
displays an asymmetric feature due to the interlayer hybridization, resulting in an effective vertical electric field to induce strong SOC at the interface.
As for the band splitting at $\mathrm{K}$ point,
almost no hybridization between two monolayers is found and 
the contribution is dominated  by $d_{xy/x^2-y^2}$ orbitals of Mo atoms, 
as shown in FIG.\ref{proj_band_sum} (IV) and (VIII).

By zooming in the two highest valence bands in FIG.\ref{resolved_spintxt} (I) 
(marked by the black rectangular box in FIG.\ref{proj_band_sum} (III)), we note that our results around $\mathrm{\Gamma}$ point 
exhibit close resemblance to the Rashba-type spin splitting in semiconductor quantum wells and surfaces of heavy
metals.\cite{Nitta:1997,LaShell:1996,Koroteev:2004,Hirahara:2006,Ir:2012,Bi_Ag111:2007,Bi_Si111:2009}
To confirm that the observed band splitting is Rashba-type, we calculate the spin components
$s_x$, $s_y$ and $s_z$ at
a constant energy cut, indicated by the blue dotted horizontal line in FIG.\ref{resolved_spintxt} (I).
As illustrated in FIG.\ref{resolved_spintxt} (II)-(IV), the spin polarizations are mainly in-plane:
$s_x$ and $s_y$ are much larger than $s_z$. Because the blue dashed line is crossing the upper splitting band, 
the inner contours show the same helical spin polarization as the outer contour, 
as shown in FIG.\ref{resolved_spintxt} (II) and (III).
These spin projections display the iconic features of the Rashba effect.
The small out-of-plane spin component $s_z$ in FIG.\ref{resolved_spintxt} (IV) is owing to the small in-plane 
potential gradient, which is also responsible for the hexagonal shape of the outer contour.\cite{LaShell:1996,Koroteev:2004}

As one can find in FIG.\ref{resolved_spintxt} (I), there are
two key properties at VBE around $\Gamma$ point: (1) the shape of the
band is very close to parabolic; (2) the spin texture is almost that given by original Rashba Hamiltonian.
Therefore, a generalized Rashba Hamiltonian is proposed to capture the physics for energies 
close to VBE around $\Gamma$ point (from about $-$0.1 eV to 0 eV),
as indicated by the red dashed line in FIG.\ref{resolved_spintxt} (I),
\begin{equation}\label{GRashbaH}
H(k_{||})=-\frac{\hbar^2 k_{||}^2}{2m}+c k_{||}+\alpha_R \vec{\sigma}\cdot(\vec{k}_{||} \times \vec{z}),
\end{equation}
where the first two terms gives rise to the so-called "sombrero hat" dispersion\cite{Rybkovskiy2014} (see the purple line around $\Gamma$
in FIG.\ref{proj_band_sum}(III)). Here, $c$ and $\alpha_R$ characterize the strength of 
the spin-independent interaction with crystal field and Rashba-type SOC, respectively.
Based on Eq.\ref{GRashbaH} the states close to VBE are suitable for the application of SFET (see the supplement).
We introduce a generalized Rashba constant $\eta_R$=$c$+$\alpha_R$ which determines the differential phase shift\cite{Datta:1990}. 
Importantly, $\eta_R$ can still be estimated by $\eta_R$=$2E_R/k_0$ with $E_R$ the generalized Rashba energy and $k_0$ the momentum offset.
From FIG.\ref{resolved_spintxt} (I), we obtain $\eta_R$=1.3 eV$\cdot$\text{\AA} by reading $E_R$=150 meV and $k_0$=0.23 \text{\AA}$^{-1}$.
The obtained $\eta_R$ for PtSe$_2$/MoSe$_2$ heterostructure is one of the largest among the Rashba constants
observed in previous studies.\cite{Nitta:1997,LaShell:1996,Koroteev:2004,Hirahara:2006,Ir:2012,Bi_Ag111:2007,Bi_Si111:2009}
Moreover, the spin splitting energy at the momentum offset  $k_0$ of VBM around $\Gamma$ point is giant, as large as 110 meV.
In addition, by calculating the whole family of PtX$_2$/MX$_2$ (M=Mo, W; X=S, Se, Te) heterostructures, we find the PtSe$_2$/MoSe$_2$ is
the most promising candidate for SFET application (see the supplement).

\begin{figure}[t!]
\centering \includegraphics[width=0.85\columnwidth]{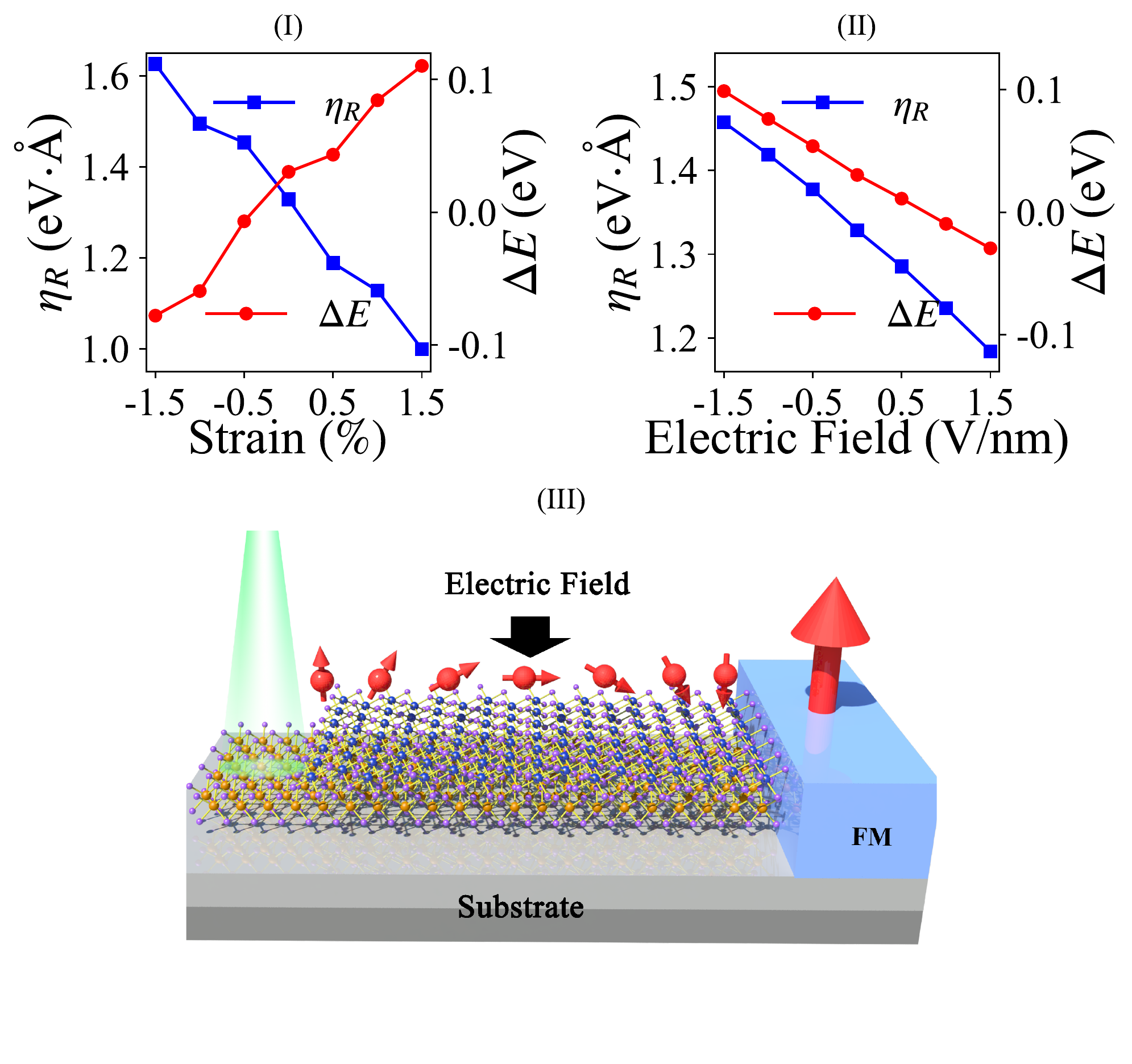}
\caption{\label{st_ef} (I-II) The generalized Rashba constant $\eta_R$ and
the relative energy shift $\Delta E$ versus the biaxial strains and external electric fields.
(III) The schematic illustration of proposed SFET.}
\end{figure}

Next, we investigate the tunability of electronic properties of PtSe$_2$/MoSe$_2$ heterostructure under biaxial strain and 
external electric field. At first, we impose biaxial strains from $\delta$=$-$1.5\% (compression) to $+1.5\%$ (tension) on the heterostructure, 
in which the strength of the strain is defined as $\delta=(a-a_0)/a_0$ with $a_0$ the original lattice constant.
For different strains, the electronic structures are calculated after all atoms are fully relaxed. 
In particular, we investigate the generalized Rashba constant $\eta_R$ and the relative energy shift 
$\Delta{E}=E^{\Gamma}_{VBM}-E^{K}_{VBM}$ between the VBMs around $\mathrm{\Gamma}$ point
and at $\mathrm{K}$ point, as shown in FIG.\ref{st_ef} (I). 
It is found that the generalized Rashba constant $\eta_R$ (in blue squares) is changed from
1.62 eV$\cdot$\text{\AA}\ to 1.00 eV$\cdot$\text{\AA} as $\delta$ is changed from
$-$1.5\% to +1.5\%. In the meanwhile, $\Delta{E}$ can be increased to $+0.1$ eV by applying $+1.5\% $ tensile strain. 
Thus, imposing tensile strain provides an efficient approach to generate an available energy interval lived by only Rashba-type states.
When applying an external electric field from $-$1.5 V/nm to 1.5 V/nm along $\vec{z}$ direction, the $\eta_R$ and $\Delta{E}$ 
are reduced at the same time, as shown in FIG.\ref{st_ef} (II), which means the external electric field also can promote the
VBM near $\mathrm{\Gamma}$ point above that at $\mathrm{K}$ point.
Furthermore, the generalized Rashba constant $\eta_R$ exhibits a notable change (24\%) from 1.46 eV$\cdot$\text{\AA} to 1.18 eV$\cdot$\text{\AA}
when the external electric field is changed from $-$1.5 V/nm to 1.5 V/nm, indicating that the spin precession can be effectively tuned with a gate voltage.
Since the difference of spin precession angle is given by\cite{Datta:1990,All_electric:2015} $\Delta{\theta}=2m\Delta{\eta_R}L/\hbar^2$ 
($\Delta{\eta_R}$ represents the change of the generalized Rashba constant, and $L$ the distance between spin injector and detector),
the appreciable tunability of $\eta_R$ is crucial for SFET application.
Note that the spin precession angle is independent of energy,\cite{Datta:1990} 
the upper split band (from $-$0.1 eV to 0 eV) around $\Gamma$ point can be utilized to realize the SFET with a light hole doping. 
The competing states from $K$ point for the upper split band can be reduced by tensile strain.

Finally, based on the tunable giant Rashba-type spin splitting
in PtSe$_2$/MoSe$_2$ heterostructure and the valley physics from H-phase monolayer MoSe$_2$,
we propose a SFET with PtSe$_2$/MoSe$_2$ as illustrated in FIG.\ref{st_ef} (III).
Similar to the SFET designed by Luo \textit{et al.},\cite{Luo:2017}
the valley polarization at $\mathrm{K}/\mathrm{K}'$ points in MoSe$_2$ is utilized
to achieve spin injection, the spin manipulation is achieved  with Rashba SOC in PtSe$_2$/MoSe$_2$
and finally a ferromagnetic contact is used to detect the spin. We estimate the minimum channel length 
(7.1 nm) of the SFET with two different electric fields ($\pm$1.5 V/nm) under $1.5\%$ tensile strain.
\footnote{The change of the effective mass $m$ is minor, we use the average value $m=0.81 m_e$ in our estimation.
Since monolayer PtSe$_2$ \cite{Wang:2015} and MoSe$_2$ \cite{Chang:2014} have already been exfoliated and proven 
to be air-stable at room temperature, it should be feasible to prepare PtSe$_2$/MoSe$_2$ heterostructure in experiment.}

In summary, we investigate the electronic properties of the PtSe$_2$/MoSe$_2$ heterostructure through
density functional theory calculations. A giant Rashba-type spin splitting
(110 meV at the momentum offset $k_0$=0.23 \AA$^{-1}$ around $\mathrm{\Gamma}$ point)
near Fermi level is predicted, arising from the emerging strong interfacial SOC.
A generalized Rashba Hamiltonian is introduced to describe the Rashba-type physics at VBE around $\Gamma$ point.
We demonstrate that both biaxial strain and external electrical field can be utilized to achieve a 2D Rashba electron system.
Furthermore, the generalized Rashba constant
$\eta_R$ in PtSe$_2$/MoSe$_2$ is as large as 1.3 eV$\cdot\text{\AA}$, and can be tuned effectively by the external electrical field.
Combining the tunable giant Rashba-type spin splitting around $\mathrm{\Gamma}$ point
in PtSe$_2$/MoSe$_2$ and the valley-physics from H-phase monolayer MoSe$_2$, we propose
a potential application for SFET.

\section*{supplementary material}
The supplementary material contains four sections: (I) the structure information of all heterostructures;
(II) the generalized Rashba Hamiltonian; (III) the orbital hybridization properties of different heterostructures; (IV) the estimation of channel length.

\begin{acknowledgments}
Q.Z. thanks the financial support from NSFC with grant No.11704121, Y.K. thanks the support from NSFC with grant No.11874265 and ShanghaiTech start-up. The authors thank HPC platform of ShanghaiTech University for providing computational facility.
\end{acknowledgments}

\bibliography{ref}

\begin{thebibliography}{44}%
\makeatletter
\providecommand \@ifxundefined [1]{%
 \@ifx{#1\undefined}
}%
\providecommand \@ifnum [1]{%
 \ifnum #1\expandafter \@firstoftwo
 \else \expandafter \@secondoftwo
 \fi
}%
\providecommand \@ifx [1]{%
 \ifx #1\expandafter \@firstoftwo
 \else \expandafter \@secondoftwo
 \fi
}%
\providecommand \natexlab [1]{#1}%
\providecommand \enquote  [1]{``#1''}%
\providecommand \bibnamefont  [1]{#1}%
\providecommand \bibfnamefont [1]{#1}%
\providecommand \citenamefont [1]{#1}%
\providecommand \href@noop [0]{\@secondoftwo}%
\providecommand \href [0]{\begingroup \@sanitize@url \@href}%
\providecommand \@href[1]{\@@startlink{#1}\@@href}%
\providecommand \@@href[1]{\endgroup#1\@@endlink}%
\providecommand \@sanitize@url [0]{\catcode `\\12\catcode `\$12\catcode
  `\&12\catcode `\#12\catcode `\^12\catcode `\_12\catcode `\%12\relax}%
\providecommand \@@startlink[1]{}%
\providecommand \@@endlink[0]{}%
\providecommand \url  [0]{\begingroup\@sanitize@url \@url }%
\providecommand \@url [1]{\endgroup\@href {#1}{\urlprefix }}%
\providecommand \urlprefix  [0]{URL }%
\providecommand \Eprint [0]{\href }%
\providecommand \doibase [0]{http://dx.doi.org/}%
\providecommand \selectlanguage [0]{\@gobble}%
\providecommand \bibinfo  [0]{\@secondoftwo}%
\providecommand \bibfield  [0]{\@secondoftwo}%
\providecommand \translation [1]{[#1]}%
\providecommand \BibitemOpen [0]{}%
\providecommand \bibitemStop [0]{}%
\providecommand \bibitemNoStop [0]{.\EOS\space}%
\providecommand \EOS [0]{\spacefactor3000\relax}%
\providecommand \BibitemShut  [1]{\csname bibitem#1\endcsname}%
\let\auto@bib@innerbib\@empty
\bibitem [{\citenamefont {Manchon}\ \emph {et~al.}(2015)\citenamefont
  {Manchon}, \citenamefont {Koo}, \citenamefont {Nitta}, \citenamefont
  {Frolov},\ and\ \citenamefont {Duine}}]{Manchon:2015}%
  \BibitemOpen
  \bibfield  {author} {\bibinfo {author} {\bibfnamefont {A.}~\bibnamefont
  {Manchon}}, \bibinfo {author} {\bibfnamefont {H.~C.}\ \bibnamefont {Koo}},
  \bibinfo {author} {\bibfnamefont {J.}~\bibnamefont {Nitta}}, \bibinfo
  {author} {\bibfnamefont {S.}~\bibnamefont {Frolov}}, \ and\ \bibinfo {author}
  {\bibfnamefont {R.}~\bibnamefont {Duine}},\ }\bibfield  {title} {\enquote
  {\bibinfo {title} {New perspectives for {Rashba} spin-orbit coupling},}\
  }\href@noop {} {\bibfield  {journal} {\bibinfo  {journal} {Nature materials}\
  }\textbf {\bibinfo {volume} {14}},\ \bibinfo {pages} {871} (\bibinfo {year}
  {2015})}\BibitemShut {NoStop}%
\bibitem [{\citenamefont {Datta}\ and\ \citenamefont {Das}(1990)}]{Datta:1990}%
  \BibitemOpen
  \bibfield  {author} {\bibinfo {author} {\bibfnamefont {S.}~\bibnamefont
  {Datta}}\ and\ \bibinfo {author} {\bibfnamefont {B.}~\bibnamefont {Das}},\
  }\bibfield  {title} {\enquote {\bibinfo {title} {Electronic analog of the
  electro‐optic modulator},}\ }\href@noop {} {\bibfield  {journal} {\bibinfo
  {journal} {Applied Physics Letters}\ }\textbf {\bibinfo {volume} {56}},\
  \bibinfo {pages} {665--667} (\bibinfo {year} {1990})}\BibitemShut {NoStop}%
\bibitem [{\citenamefont {Bychkov}\ and\ \citenamefont
  {Rashba}(1984)}]{Rashba:1984}%
  \BibitemOpen
  \bibfield  {author} {\bibinfo {author} {\bibfnamefont {Y.~A.}\ \bibnamefont
  {Bychkov}}\ and\ \bibinfo {author} {\bibfnamefont {{\'E}.~I.}\ \bibnamefont
  {Rashba}},\ }\bibfield  {title} {\enquote {\bibinfo {title} {Properties of a
  {2D} electron gas with lifted spectral degeneracy},}\ }\href@noop {}
  {\bibfield  {journal} {\bibinfo  {journal} {JETP lett}\ }\textbf {\bibinfo
  {volume} {39}},\ \bibinfo {pages} {78} (\bibinfo {year} {1984})}\BibitemShut
  {NoStop}%
\bibitem [{\citenamefont {Nitta}\ \emph {et~al.}(1997)\citenamefont {Nitta},
  \citenamefont {Akazaki}, \citenamefont {Takayanagi},\ and\ \citenamefont
  {Enoki}}]{Nitta:1997}%
  \BibitemOpen
  \bibfield  {author} {\bibinfo {author} {\bibfnamefont {J.}~\bibnamefont
  {Nitta}}, \bibinfo {author} {\bibfnamefont {T.}~\bibnamefont {Akazaki}},
  \bibinfo {author} {\bibfnamefont {H.}~\bibnamefont {Takayanagi}}, \ and\
  \bibinfo {author} {\bibfnamefont {T.}~\bibnamefont {Enoki}},\ }\bibfield
  {title} {\enquote {\bibinfo {title} {Gate control of spin-orbit interaction
  in an inverted {In}$_{0.53}${Ga}$_{0.47}${As}/{In}$_{0.52}${Al}$_{0.48}${As}
  heterostructure},}\ }\href@noop {} {\bibfield  {journal} {\bibinfo  {journal}
  {Phys. Rev. Lett.}\ }\textbf {\bibinfo {volume} {78}},\ \bibinfo {pages}
  {1335--1338} (\bibinfo {year} {1997})}\BibitemShut {NoStop}%
\bibitem [{\citenamefont {LaShell}, \citenamefont {McDougall},\ and\
  \citenamefont {Jensen}(1996)}]{LaShell:1996}%
  \BibitemOpen
  \bibfield  {author} {\bibinfo {author} {\bibfnamefont {S.}~\bibnamefont
  {LaShell}}, \bibinfo {author} {\bibfnamefont {B.~A.}\ \bibnamefont
  {McDougall}}, \ and\ \bibinfo {author} {\bibfnamefont {E.}~\bibnamefont
  {Jensen}},\ }\bibfield  {title} {\enquote {\bibinfo {title} {Spin splitting
  of an {Au}(111) surface state band observed with angle resolved photoelectron
  spectroscopy},}\ }\href@noop {} {\bibfield  {journal} {\bibinfo  {journal}
  {Phys. Rev. Lett.}\ }\textbf {\bibinfo {volume} {77}},\ \bibinfo {pages}
  {3419--3422} (\bibinfo {year} {1996})}\BibitemShut {NoStop}%
\bibitem [{\citenamefont {Koroteev}\ \emph {et~al.}(2004)\citenamefont
  {Koroteev}, \citenamefont {Bihlmayer}, \citenamefont {Gayone}, \citenamefont
  {Chulkov}, \citenamefont {Bl\"ugel}, \citenamefont {Echenique},\ and\
  \citenamefont {Hofmann}}]{Koroteev:2004}%
  \BibitemOpen
  \bibfield  {author} {\bibinfo {author} {\bibfnamefont {Y.~M.}\ \bibnamefont
  {Koroteev}}, \bibinfo {author} {\bibfnamefont {G.}~\bibnamefont {Bihlmayer}},
  \bibinfo {author} {\bibfnamefont {J.~E.}\ \bibnamefont {Gayone}}, \bibinfo
  {author} {\bibfnamefont {E.~V.}\ \bibnamefont {Chulkov}}, \bibinfo {author}
  {\bibfnamefont {S.}~\bibnamefont {Bl\"ugel}}, \bibinfo {author}
  {\bibfnamefont {P.~M.}\ \bibnamefont {Echenique}}, \ and\ \bibinfo {author}
  {\bibfnamefont {P.}~\bibnamefont {Hofmann}},\ }\bibfield  {title} {\enquote
  {\bibinfo {title} {Strong spin-orbit splitting on {Bi} surfaces},}\
  }\href@noop {} {\bibfield  {journal} {\bibinfo  {journal} {Phys. Rev. Lett.}\
  }\textbf {\bibinfo {volume} {93}},\ \bibinfo {pages} {046403} (\bibinfo
  {year} {2004})}\BibitemShut {NoStop}%
\bibitem [{\citenamefont {Hirahara}\ \emph {et~al.}(2006)\citenamefont
  {Hirahara}, \citenamefont {Nagao}, \citenamefont {Matsuda}, \citenamefont
  {Bihlmayer}, \citenamefont {Chulkov}, \citenamefont {Koroteev}, \citenamefont
  {Echenique}, \citenamefont {Saito},\ and\ \citenamefont
  {Hasegawa}}]{Hirahara:2006}%
  \BibitemOpen
  \bibfield  {author} {\bibinfo {author} {\bibfnamefont {T.}~\bibnamefont
  {Hirahara}}, \bibinfo {author} {\bibfnamefont {T.}~\bibnamefont {Nagao}},
  \bibinfo {author} {\bibfnamefont {I.}~\bibnamefont {Matsuda}}, \bibinfo
  {author} {\bibfnamefont {G.}~\bibnamefont {Bihlmayer}}, \bibinfo {author}
  {\bibfnamefont {E.~V.}\ \bibnamefont {Chulkov}}, \bibinfo {author}
  {\bibfnamefont {Y.~M.}\ \bibnamefont {Koroteev}}, \bibinfo {author}
  {\bibfnamefont {P.~M.}\ \bibnamefont {Echenique}}, \bibinfo {author}
  {\bibfnamefont {M.}~\bibnamefont {Saito}}, \ and\ \bibinfo {author}
  {\bibfnamefont {S.}~\bibnamefont {Hasegawa}},\ }\bibfield  {title} {\enquote
  {\bibinfo {title} {Role of spin-orbit coupling and hybridization effects in
  the electronic structure of ultrathin {Bi} films},}\ }\href@noop {}
  {\bibfield  {journal} {\bibinfo  {journal} {Phys. Rev. Lett.}\ }\textbf
  {\bibinfo {volume} {97}},\ \bibinfo {pages} {146803} (\bibinfo {year}
  {2006})}\BibitemShut {NoStop}%
\bibitem [{\citenamefont {Varykhalov}\ \emph {et~al.}(2012)\citenamefont
  {Varykhalov}, \citenamefont {Marchenko}, \citenamefont {Scholz},
  \citenamefont {Rienks}, \citenamefont {Kim}, \citenamefont {Bihlmayer},
  \citenamefont {S\'anchez-Barriga},\ and\ \citenamefont {Rader}}]{Ir:2012}%
  \BibitemOpen
  \bibfield  {author} {\bibinfo {author} {\bibfnamefont {A.}~\bibnamefont
  {Varykhalov}}, \bibinfo {author} {\bibfnamefont {D.}~\bibnamefont
  {Marchenko}}, \bibinfo {author} {\bibfnamefont {M.~R.}\ \bibnamefont
  {Scholz}}, \bibinfo {author} {\bibfnamefont {E.~D.~L.}\ \bibnamefont
  {Rienks}}, \bibinfo {author} {\bibfnamefont {T.~K.}\ \bibnamefont {Kim}},
  \bibinfo {author} {\bibfnamefont {G.}~\bibnamefont {Bihlmayer}}, \bibinfo
  {author} {\bibfnamefont {J.}~\bibnamefont {S\'anchez-Barriga}}, \ and\
  \bibinfo {author} {\bibfnamefont {O.}~\bibnamefont {Rader}},\ }\bibfield
  {title} {\enquote {\bibinfo {title} {{Ir}(111) surface state with giant
  {Rashba} splitting persists under graphene in air},}\ }\href@noop {}
  {\bibfield  {journal} {\bibinfo  {journal} {Phys. Rev. Lett.}\ }\textbf
  {\bibinfo {volume} {108}},\ \bibinfo {pages} {066804} (\bibinfo {year}
  {2012})}\BibitemShut {NoStop}%
\bibitem [{\citenamefont {Dil}\ \emph {et~al.}(2008)\citenamefont {Dil},
  \citenamefont {Meier}, \citenamefont {Lobo-Checa}, \citenamefont {Patthey},
  \citenamefont {Bihlmayer},\ and\ \citenamefont {Osterwalder}}]{Hugo:2008}%
  \BibitemOpen
  \bibfield  {author} {\bibinfo {author} {\bibfnamefont {J.~H.}\ \bibnamefont
  {Dil}}, \bibinfo {author} {\bibfnamefont {F.}~\bibnamefont {Meier}}, \bibinfo
  {author} {\bibfnamefont {J.}~\bibnamefont {Lobo-Checa}}, \bibinfo {author}
  {\bibfnamefont {L.}~\bibnamefont {Patthey}}, \bibinfo {author} {\bibfnamefont
  {G.}~\bibnamefont {Bihlmayer}}, \ and\ \bibinfo {author} {\bibfnamefont
  {J.}~\bibnamefont {Osterwalder}},\ }\bibfield  {title} {\enquote {\bibinfo
  {title} {Rashba-type spin-orbit splitting of quantum well states in ultrathin
  {Pb} films},}\ }\href@noop {} {\bibfield  {journal} {\bibinfo  {journal}
  {Phys. Rev. Lett.}\ }\textbf {\bibinfo {volume} {101}},\ \bibinfo {pages}
  {266802} (\bibinfo {year} {2008})}\BibitemShut {NoStop}%
\bibitem [{\citenamefont {King}\ \emph {et~al.}(2011)\citenamefont {King},
  \citenamefont {Hatch}, \citenamefont {Bianchi}, \citenamefont {Ovsyannikov},
  \citenamefont {Lupulescu}, \citenamefont {Landolt}, \citenamefont {Slomski},
  \citenamefont {Dil}, \citenamefont {Guan}, \citenamefont {Mi}, \citenamefont
  {Rienks}, \citenamefont {Fink}, \citenamefont {Lindblad}, \citenamefont
  {Svensson}, \citenamefont {Bao}, \citenamefont {Balakrishnan}, \citenamefont
  {Iversen}, \citenamefont {Osterwalder}, \citenamefont {Eberhardt},
  \citenamefont {Baumberger},\ and\ \citenamefont {Hofmann}}]{King:2011}%
  \BibitemOpen
  \bibfield  {author} {\bibinfo {author} {\bibfnamefont {P.~D.~C.}\
  \bibnamefont {King}}, \bibinfo {author} {\bibfnamefont {R.~C.}\ \bibnamefont
  {Hatch}}, \bibinfo {author} {\bibfnamefont {M.}~\bibnamefont {Bianchi}},
  \bibinfo {author} {\bibfnamefont {R.}~\bibnamefont {Ovsyannikov}}, \bibinfo
  {author} {\bibfnamefont {C.}~\bibnamefont {Lupulescu}}, \bibinfo {author}
  {\bibfnamefont {G.}~\bibnamefont {Landolt}}, \bibinfo {author} {\bibfnamefont
  {B.}~\bibnamefont {Slomski}}, \bibinfo {author} {\bibfnamefont {J.~H.}\
  \bibnamefont {Dil}}, \bibinfo {author} {\bibfnamefont {D.}~\bibnamefont
  {Guan}}, \bibinfo {author} {\bibfnamefont {J.~L.}\ \bibnamefont {Mi}},
  \bibinfo {author} {\bibfnamefont {E.~D.~L.}\ \bibnamefont {Rienks}}, \bibinfo
  {author} {\bibfnamefont {J.}~\bibnamefont {Fink}}, \bibinfo {author}
  {\bibfnamefont {A.}~\bibnamefont {Lindblad}}, \bibinfo {author}
  {\bibfnamefont {S.}~\bibnamefont {Svensson}}, \bibinfo {author}
  {\bibfnamefont {S.}~\bibnamefont {Bao}}, \bibinfo {author} {\bibfnamefont
  {G.}~\bibnamefont {Balakrishnan}}, \bibinfo {author} {\bibfnamefont {B.~B.}\
  \bibnamefont {Iversen}}, \bibinfo {author} {\bibfnamefont {J.}~\bibnamefont
  {Osterwalder}}, \bibinfo {author} {\bibfnamefont {W.}~\bibnamefont
  {Eberhardt}}, \bibinfo {author} {\bibfnamefont {F.}~\bibnamefont
  {Baumberger}}, \ and\ \bibinfo {author} {\bibfnamefont {P.}~\bibnamefont
  {Hofmann}},\ }\bibfield  {title} {\enquote {\bibinfo {title} {Large tunable
  {Rashba} spin splitting of a two-dimensional electron gas in
  {Bi}$_{2}${Se}$_{3}$},}\ }\href@noop {} {\bibfield  {journal} {\bibinfo
  {journal} {Phys. Rev. Lett.}\ }\textbf {\bibinfo {volume} {107}},\ \bibinfo
  {pages} {096802} (\bibinfo {year} {2011})}\BibitemShut {NoStop}%
\bibitem [{\citenamefont {Zhu}\ \emph {et~al.}(2011)\citenamefont {Zhu},
  \citenamefont {Levy}, \citenamefont {Ludbrook}, \citenamefont {Veenstra},
  \citenamefont {Rosen}, \citenamefont {Comin}, \citenamefont {Wong},
  \citenamefont {Dosanjh}, \citenamefont {Ubaldini}, \citenamefont {Syers},
  \citenamefont {Butch}, \citenamefont {Paglione}, \citenamefont {Elfimov},\
  and\ \citenamefont {Damascelli}}]{Zhu:2011}%
  \BibitemOpen
  \bibfield  {author} {\bibinfo {author} {\bibfnamefont {Z.-H.}\ \bibnamefont
  {Zhu}}, \bibinfo {author} {\bibfnamefont {G.}~\bibnamefont {Levy}}, \bibinfo
  {author} {\bibfnamefont {B.}~\bibnamefont {Ludbrook}}, \bibinfo {author}
  {\bibfnamefont {C.~N.}\ \bibnamefont {Veenstra}}, \bibinfo {author}
  {\bibfnamefont {J.~A.}\ \bibnamefont {Rosen}}, \bibinfo {author}
  {\bibfnamefont {R.}~\bibnamefont {Comin}}, \bibinfo {author} {\bibfnamefont
  {D.}~\bibnamefont {Wong}}, \bibinfo {author} {\bibfnamefont {P.}~\bibnamefont
  {Dosanjh}}, \bibinfo {author} {\bibfnamefont {A.}~\bibnamefont {Ubaldini}},
  \bibinfo {author} {\bibfnamefont {P.}~\bibnamefont {Syers}}, \bibinfo
  {author} {\bibfnamefont {N.~P.}\ \bibnamefont {Butch}}, \bibinfo {author}
  {\bibfnamefont {J.}~\bibnamefont {Paglione}}, \bibinfo {author}
  {\bibfnamefont {I.~S.}\ \bibnamefont {Elfimov}}, \ and\ \bibinfo {author}
  {\bibfnamefont {A.}~\bibnamefont {Damascelli}},\ }\bibfield  {title}
  {\enquote {\bibinfo {title} {Rashba spin-splitting control at the surface of
  the topological insulator {Bi}$_{2}${Se}$_{3}$},}\ }\href@noop {} {\bibfield
  {journal} {\bibinfo  {journal} {Phys. Rev. Lett.}\ }\textbf {\bibinfo
  {volume} {107}},\ \bibinfo {pages} {186405} (\bibinfo {year}
  {2011})}\BibitemShut {NoStop}%
\bibitem [{\citenamefont {Ast}\ \emph {et~al.}(2007)\citenamefont {Ast},
  \citenamefont {Henk}, \citenamefont {Ernst}, \citenamefont {Moreschini},
  \citenamefont {Falub}, \citenamefont {Pacil\'e}, \citenamefont {Bruno},
  \citenamefont {Kern},\ and\ \citenamefont {Grioni}}]{Bi_Ag111:2007}%
  \BibitemOpen
  \bibfield  {author} {\bibinfo {author} {\bibfnamefont {C.~R.}\ \bibnamefont
  {Ast}}, \bibinfo {author} {\bibfnamefont {J.}~\bibnamefont {Henk}}, \bibinfo
  {author} {\bibfnamefont {A.}~\bibnamefont {Ernst}}, \bibinfo {author}
  {\bibfnamefont {L.}~\bibnamefont {Moreschini}}, \bibinfo {author}
  {\bibfnamefont {M.~C.}\ \bibnamefont {Falub}}, \bibinfo {author}
  {\bibfnamefont {D.}~\bibnamefont {Pacil\'e}}, \bibinfo {author}
  {\bibfnamefont {P.}~\bibnamefont {Bruno}}, \bibinfo {author} {\bibfnamefont
  {K.}~\bibnamefont {Kern}}, \ and\ \bibinfo {author} {\bibfnamefont
  {M.}~\bibnamefont {Grioni}},\ }\bibfield  {title} {\enquote {\bibinfo {title}
  {Giant spin splitting through surface alloying},}\ }\href@noop {} {\bibfield
  {journal} {\bibinfo  {journal} {Phys. Rev. Lett.}\ }\textbf {\bibinfo
  {volume} {98}},\ \bibinfo {pages} {186807} (\bibinfo {year}
  {2007})}\BibitemShut {NoStop}%
\bibitem [{\citenamefont {Gierz}\ \emph {et~al.}(2009)\citenamefont {Gierz},
  \citenamefont {Suzuki}, \citenamefont {Frantzeskakis}, \citenamefont {Pons},
  \citenamefont {Ostanin}, \citenamefont {Ernst}, \citenamefont {Henk},
  \citenamefont {Grioni}, \citenamefont {Kern},\ and\ \citenamefont
  {Ast}}]{Bi_Si111:2009}%
  \BibitemOpen
  \bibfield  {author} {\bibinfo {author} {\bibfnamefont {I.}~\bibnamefont
  {Gierz}}, \bibinfo {author} {\bibfnamefont {T.}~\bibnamefont {Suzuki}},
  \bibinfo {author} {\bibfnamefont {E.}~\bibnamefont {Frantzeskakis}}, \bibinfo
  {author} {\bibfnamefont {S.}~\bibnamefont {Pons}}, \bibinfo {author}
  {\bibfnamefont {S.}~\bibnamefont {Ostanin}}, \bibinfo {author} {\bibfnamefont
  {A.}~\bibnamefont {Ernst}}, \bibinfo {author} {\bibfnamefont
  {J.}~\bibnamefont {Henk}}, \bibinfo {author} {\bibfnamefont {M.}~\bibnamefont
  {Grioni}}, \bibinfo {author} {\bibfnamefont {K.}~\bibnamefont {Kern}}, \ and\
  \bibinfo {author} {\bibfnamefont {C.~R.}\ \bibnamefont {Ast}},\ }\bibfield
  {title} {\enquote {\bibinfo {title} {Silicon surface with giant spin
  splitting},}\ }\href@noop {} {\bibfield  {journal} {\bibinfo  {journal}
  {Phys. Rev. Lett.}\ }\textbf {\bibinfo {volume} {103}},\ \bibinfo {pages}
  {046803} (\bibinfo {year} {2009})}\BibitemShut {NoStop}%
\bibitem [{\citenamefont {Bindel}\ \emph {et~al.}(2016)\citenamefont {Bindel},
  \citenamefont {Pezzotta}, \citenamefont {Ulrich}, \citenamefont {Liebmann},
  \citenamefont {Sherman},\ and\ \citenamefont {Morgenstern}}]{Bindel:2016}%
  \BibitemOpen
  \bibfield  {author} {\bibinfo {author} {\bibfnamefont {J.~R.}\ \bibnamefont
  {Bindel}}, \bibinfo {author} {\bibfnamefont {M.}~\bibnamefont {Pezzotta}},
  \bibinfo {author} {\bibfnamefont {J.}~\bibnamefont {Ulrich}}, \bibinfo
  {author} {\bibfnamefont {M.}~\bibnamefont {Liebmann}}, \bibinfo {author}
  {\bibfnamefont {E.~Y.}\ \bibnamefont {Sherman}}, \ and\ \bibinfo {author}
  {\bibfnamefont {M.}~\bibnamefont {Morgenstern}},\ }\bibfield  {title}
  {\enquote {\bibinfo {title} {Probing variations of the {Rashba} spin--orbit
  coupling at the nanometre scale},}\ }\href@noop {} {\bibfield  {journal}
  {\bibinfo  {journal} {Nature Physics}\ }\textbf {\bibinfo {volume} {12}},\
  \bibinfo {pages} {920} (\bibinfo {year} {2016})}\BibitemShut {NoStop}%
\bibitem [{\citenamefont {Ishizaka}\ \emph {et~al.}(2011)\citenamefont
  {Ishizaka}, \citenamefont {Bahramy}, \citenamefont {Murakawa}, \citenamefont
  {Sakano}, \citenamefont {Shimojima}, \citenamefont {Sonobe}, \citenamefont
  {Koizumi}, \citenamefont {Shin}, \citenamefont {Miyahara}, \citenamefont
  {Kimura}, \citenamefont {Miyamoto}, \citenamefont {Okuda}, \citenamefont
  {Namatame}, \citenamefont {Taniguchi}, \citenamefont {Arita}, \citenamefont
  {Nagaosa}, \citenamefont {Kobayashi}, \citenamefont {Murakami}, \citenamefont
  {Kumai}, \citenamefont {Kaneko}, \citenamefont {Onose},\ and\ \citenamefont
  {Tokura}}]{Ishizaka:2011}%
  \BibitemOpen
  \bibfield  {author} {\bibinfo {author} {\bibfnamefont {K.}~\bibnamefont
  {Ishizaka}}, \bibinfo {author} {\bibfnamefont {M.~S.}\ \bibnamefont
  {Bahramy}}, \bibinfo {author} {\bibfnamefont {H.}~\bibnamefont {Murakawa}},
  \bibinfo {author} {\bibfnamefont {M.}~\bibnamefont {Sakano}}, \bibinfo
  {author} {\bibfnamefont {T.}~\bibnamefont {Shimojima}}, \bibinfo {author}
  {\bibfnamefont {T.}~\bibnamefont {Sonobe}}, \bibinfo {author} {\bibfnamefont
  {K.}~\bibnamefont {Koizumi}}, \bibinfo {author} {\bibfnamefont
  {S.}~\bibnamefont {Shin}}, \bibinfo {author} {\bibfnamefont {H.}~\bibnamefont
  {Miyahara}}, \bibinfo {author} {\bibfnamefont {A.}~\bibnamefont {Kimura}},
  \bibinfo {author} {\bibfnamefont {K.}~\bibnamefont {Miyamoto}}, \bibinfo
  {author} {\bibfnamefont {T.}~\bibnamefont {Okuda}}, \bibinfo {author}
  {\bibfnamefont {H.}~\bibnamefont {Namatame}}, \bibinfo {author}
  {\bibfnamefont {M.}~\bibnamefont {Taniguchi}}, \bibinfo {author}
  {\bibfnamefont {R.}~\bibnamefont {Arita}}, \bibinfo {author} {\bibfnamefont
  {N.}~\bibnamefont {Nagaosa}}, \bibinfo {author} {\bibfnamefont
  {K.}~\bibnamefont {Kobayashi}}, \bibinfo {author} {\bibfnamefont
  {Y.}~\bibnamefont {Murakami}}, \bibinfo {author} {\bibfnamefont
  {R.}~\bibnamefont {Kumai}}, \bibinfo {author} {\bibfnamefont
  {Y.}~\bibnamefont {Kaneko}}, \bibinfo {author} {\bibfnamefont
  {Y.}~\bibnamefont {Onose}}, \ and\ \bibinfo {author} {\bibfnamefont
  {Y.}~\bibnamefont {Tokura}},\ }\bibfield  {title} {\enquote {\bibinfo {title}
  {Giant {Rashba}-type spin splitting in bulk {BiTeI}},}\ }\href@noop {}
  {\bibfield  {journal} {\bibinfo  {journal} {Nature Materials}\ }\textbf
  {\bibinfo {volume} {10}},\ \bibinfo {pages} {521--526} (\bibinfo {year}
  {2011})}\BibitemShut {NoStop}%
\bibitem [{\citenamefont {Di~Sante}\ \emph {et~al.}(2013)\citenamefont
  {Di~Sante}, \citenamefont {Barone}, \citenamefont {Bertacco},\ and\
  \citenamefont {Picozzi}}]{Sante:2013}%
  \BibitemOpen
  \bibfield  {author} {\bibinfo {author} {\bibfnamefont {D.}~\bibnamefont
  {Di~Sante}}, \bibinfo {author} {\bibfnamefont {P.}~\bibnamefont {Barone}},
  \bibinfo {author} {\bibfnamefont {R.}~\bibnamefont {Bertacco}}, \ and\
  \bibinfo {author} {\bibfnamefont {S.}~\bibnamefont {Picozzi}},\ }\bibfield
  {title} {\enquote {\bibinfo {title} {Electric control of the giant {Rashba}
  effect in bulk {GeTe}},}\ }\href@noop {} {\bibfield  {journal} {\bibinfo
  {journal} {Advanced Materials}\ }\textbf {\bibinfo {volume} {25}},\ \bibinfo
  {pages} {509--513} (\bibinfo {year} {2013})}\BibitemShut {NoStop}%
\bibitem [{\citenamefont {Liebmann}\ \emph {et~al.}(2016)\citenamefont
  {Liebmann}, \citenamefont {Rinaldi}, \citenamefont {Di~Sante}, \citenamefont
  {Kellner}, \citenamefont {Pauly}, \citenamefont {Wang}, \citenamefont
  {Boschker}, \citenamefont {Giussani}, \citenamefont {Bertoli}, \citenamefont
  {Cantoni} \emph {et~al.}}]{Liebmann:2016}%
  \BibitemOpen
  \bibfield  {author} {\bibinfo {author} {\bibfnamefont {M.}~\bibnamefont
  {Liebmann}}, \bibinfo {author} {\bibfnamefont {C.}~\bibnamefont {Rinaldi}},
  \bibinfo {author} {\bibfnamefont {D.}~\bibnamefont {Di~Sante}}, \bibinfo
  {author} {\bibfnamefont {J.}~\bibnamefont {Kellner}}, \bibinfo {author}
  {\bibfnamefont {C.}~\bibnamefont {Pauly}}, \bibinfo {author} {\bibfnamefont
  {R.~N.}\ \bibnamefont {Wang}}, \bibinfo {author} {\bibfnamefont {J.~E.}\
  \bibnamefont {Boschker}}, \bibinfo {author} {\bibfnamefont {A.}~\bibnamefont
  {Giussani}}, \bibinfo {author} {\bibfnamefont {S.}~\bibnamefont {Bertoli}},
  \bibinfo {author} {\bibfnamefont {M.}~\bibnamefont {Cantoni}},  \emph
  {et~al.},\ }\bibfield  {title} {\enquote {\bibinfo {title} {Giant
  {Rashba}-type spin splitting in ferroelectric {GeTe} (111)},}\ }\href@noop {}
  {\bibfield  {journal} {\bibinfo  {journal} {Advanced Materials}\ }\textbf
  {\bibinfo {volume} {28}},\ \bibinfo {pages} {560--565} (\bibinfo {year}
  {2016})}\BibitemShut {NoStop}%
\bibitem [{\citenamefont {Ma}\ \emph {et~al.}(2014)\citenamefont {Ma},
  \citenamefont {Dai}, \citenamefont {Wei}, \citenamefont {Li},\ and\
  \citenamefont {Huang}}]{Ma:2014}%
  \BibitemOpen
  \bibfield  {author} {\bibinfo {author} {\bibfnamefont {Y.}~\bibnamefont
  {Ma}}, \bibinfo {author} {\bibfnamefont {Y.}~\bibnamefont {Dai}}, \bibinfo
  {author} {\bibfnamefont {W.}~\bibnamefont {Wei}}, \bibinfo {author}
  {\bibfnamefont {X.}~\bibnamefont {Li}}, \ and\ \bibinfo {author}
  {\bibfnamefont {B.}~\bibnamefont {Huang}},\ }\bibfield  {title} {\enquote
  {\bibinfo {title} {Emergence of electric polarity in {BiTeX} ({X}= {Br} and
  {I}) monolayers and the giant {Rashba} spin splitting},}\ }\href@noop {}
  {\bibfield  {journal} {\bibinfo  {journal} {Physical Chemistry Chemical
  Physics}\ }\textbf {\bibinfo {volume} {16}},\ \bibinfo {pages} {17603--17609}
  (\bibinfo {year} {2014})}\BibitemShut {NoStop}%
\bibitem [{\citenamefont {Ajayan}, \citenamefont {Kim},\ and\ \citenamefont
  {Banerjee}(2016)}]{Ajayan:2016}%
  \BibitemOpen
  \bibfield  {author} {\bibinfo {author} {\bibfnamefont {P.}~\bibnamefont
  {Ajayan}}, \bibinfo {author} {\bibfnamefont {P.}~\bibnamefont {Kim}}, \ and\
  \bibinfo {author} {\bibfnamefont {K.}~\bibnamefont {Banerjee}},\ }\bibfield
  {title} {\enquote {\bibinfo {title} {Two-dimensional van der {Waals}
  materials},}\ }\href@noop {} {\bibfield  {journal} {\bibinfo  {journal}
  {Physics Today}\ }\textbf {\bibinfo {volume} {69}},\ \bibinfo {pages}
  {38--44} (\bibinfo {year} {2016})}\BibitemShut {NoStop}%
\bibitem [{\citenamefont {Novoselov}\ \emph {et~al.}(2016)\citenamefont
  {Novoselov}, \citenamefont {Mishchenko}, \citenamefont {Carvalho},\ and\
  \citenamefont {Castro~Neto}}]{Novoselov:2016}%
  \BibitemOpen
  \bibfield  {author} {\bibinfo {author} {\bibfnamefont {K.~S.}\ \bibnamefont
  {Novoselov}}, \bibinfo {author} {\bibfnamefont {A.}~\bibnamefont
  {Mishchenko}}, \bibinfo {author} {\bibfnamefont {A.}~\bibnamefont
  {Carvalho}}, \ and\ \bibinfo {author} {\bibfnamefont {A.~H.}\ \bibnamefont
  {Castro~Neto}},\ }\bibfield  {title} {\enquote {\bibinfo {title} {{2D}
  materials and van der {Waals} heterostructures},}\ }\href@noop {} {\bibfield
  {journal} {\bibinfo  {journal} {Science}\ }\textbf {\bibinfo {volume}
  {353}},\ \bibinfo {pages} {aac9439} (\bibinfo {year} {2016})}\BibitemShut
  {NoStop}%
\bibitem [{\citenamefont {Kasai}\ \emph {et~al.}(2018)\citenamefont {Kasai},
  \citenamefont {Tolborg}, \citenamefont {Sist}, \citenamefont {Zhang},
  \citenamefont {Hathwar}, \citenamefont {Fils{\o}}, \citenamefont {Cenedese},
  \citenamefont {Sugimoto}, \citenamefont {Overgaard}, \citenamefont
  {Nishibori} \emph {et~al.}}]{Kasai:2018}%
  \BibitemOpen
  \bibfield  {author} {\bibinfo {author} {\bibfnamefont {H.}~\bibnamefont
  {Kasai}}, \bibinfo {author} {\bibfnamefont {K.}~\bibnamefont {Tolborg}},
  \bibinfo {author} {\bibfnamefont {M.}~\bibnamefont {Sist}}, \bibinfo {author}
  {\bibfnamefont {J.}~\bibnamefont {Zhang}}, \bibinfo {author} {\bibfnamefont
  {V.~R.}\ \bibnamefont {Hathwar}}, \bibinfo {author} {\bibfnamefont
  {M.~{\O}.}\ \bibnamefont {Fils{\o}}}, \bibinfo {author} {\bibfnamefont
  {S.}~\bibnamefont {Cenedese}}, \bibinfo {author} {\bibfnamefont
  {K.}~\bibnamefont {Sugimoto}}, \bibinfo {author} {\bibfnamefont
  {J.}~\bibnamefont {Overgaard}}, \bibinfo {author} {\bibfnamefont
  {E.}~\bibnamefont {Nishibori}},  \emph {et~al.},\ }\bibfield  {title}
  {\enquote {\bibinfo {title} {X-ray electron density investigation of chemical
  bonding in van der {Waals} materials},}\ }\href@noop {} {\bibfield  {journal}
  {\bibinfo  {journal} {Nature materials}\ }\textbf {\bibinfo {volume} {17}},\
  \bibinfo {pages} {249} (\bibinfo {year} {2018})}\BibitemShut {NoStop}%
\bibitem [{\citenamefont {Liu}, \citenamefont {Guo},\ and\ \citenamefont
  {Freeman}(2013)}]{Liuqihang:2013}%
  \BibitemOpen
  \bibfield  {author} {\bibinfo {author} {\bibfnamefont {Q.}~\bibnamefont
  {Liu}}, \bibinfo {author} {\bibfnamefont {Y.}~\bibnamefont {Guo}}, \ and\
  \bibinfo {author} {\bibfnamefont {A.~J.}\ \bibnamefont {Freeman}},\
  }\bibfield  {title} {\enquote {\bibinfo {title} {Tunable {Rashba} effect in
  two-dimensional {LaOBiS}$_2$ films: Ultrathin candidates for spin field
  effect transistors},}\ }\href@noop {} {\bibfield  {journal} {\bibinfo
  {journal} {Nano letters}\ }\textbf {\bibinfo {volume} {13}},\ \bibinfo
  {pages} {5264--5270} (\bibinfo {year} {2013})}\BibitemShut {NoStop}%
\bibitem [{\citenamefont {Singh}\ and\ \citenamefont
  {Romero}(2017)}]{Singh:2017}%
  \BibitemOpen
  \bibfield  {author} {\bibinfo {author} {\bibfnamefont {S.}~\bibnamefont
  {Singh}}\ and\ \bibinfo {author} {\bibfnamefont {A.~H.}\ \bibnamefont
  {Romero}},\ }\bibfield  {title} {\enquote {\bibinfo {title} {Giant tunable
  rashba spin splitting in a two-dimensional {BiSb} monolayer and in
  {BiSb}/{AlN} heterostructures},}\ }\href@noop {} {\bibfield  {journal}
  {\bibinfo  {journal} {Phys. Rev. B}\ }\textbf {\bibinfo {volume} {95}},\
  \bibinfo {pages} {165444} (\bibinfo {year} {2017})}\BibitemShut {NoStop}%
\bibitem [{\citenamefont {Wang}\ and\ \citenamefont {Jeng}(2017)}]{Wang:2017}%
  \BibitemOpen
  \bibfield  {author} {\bibinfo {author} {\bibfnamefont {T.-H.}\ \bibnamefont
  {Wang}}\ and\ \bibinfo {author} {\bibfnamefont {H.-T.}\ \bibnamefont
  {Jeng}},\ }\bibfield  {title} {\enquote {\bibinfo {title} {Wide-range ideal
  {2D} {Rashba} electron gas with large spin splitting in
  {Bi}$_2${Se}$_3$/{MoTe}$_2$ heterostructure},}\ }\href@noop {} {\bibfield
  {journal} {\bibinfo  {journal} {npj Comput. Mater.}\ }\textbf {\bibinfo
  {volume} {3}},\ \bibinfo {pages} {5--10} (\bibinfo {year}
  {2017})}\BibitemShut {NoStop}%
\bibitem [{\citenamefont {Zhang}\ and\ \citenamefont
  {Schwingenschl\"ogl}(2018)}]{Zhang:2018}%
  \BibitemOpen
  \bibfield  {author} {\bibinfo {author} {\bibfnamefont {Q.}~\bibnamefont
  {Zhang}}\ and\ \bibinfo {author} {\bibfnamefont {U.}~\bibnamefont
  {Schwingenschl\"ogl}},\ }\bibfield  {title} {\enquote {\bibinfo {title}
  {Rashba effect and enriched spin-valley coupling in {GaX}/{MX}$_2$
  ({M}={Mo},{W}; {X}={S},{Se},{Te}) heterostructures},}\ }\href@noop {}
  {\bibfield  {journal} {\bibinfo  {journal} {Phys. Rev. B}\ }\textbf {\bibinfo
  {volume} {97}},\ \bibinfo {pages} {155415} (\bibinfo {year}
  {2018})}\BibitemShut {NoStop}%
\bibitem [{\citenamefont {Wang}\ \emph {et~al.}(2015)\citenamefont {Wang},
  \citenamefont {Li}, \citenamefont {Yao}, \citenamefont {Song}, \citenamefont
  {Sun}, \citenamefont {Pan}, \citenamefont {Ren}, \citenamefont {Li},
  \citenamefont {Okunishi}, \citenamefont {Wang}, \citenamefont {Wang},
  \citenamefont {Shao}, \citenamefont {Zhang}, \citenamefont {Yang},
  \citenamefont {Schwier}, \citenamefont {Iwasawa}, \citenamefont {Shimada},
  \citenamefont {Taniguchi}, \citenamefont {Cheng}, \citenamefont {Zhou},
  \citenamefont {Du}, \citenamefont {Pennycook}, \citenamefont {Pantelides},\
  and\ \citenamefont {Gao}}]{Wang:2015}%
  \BibitemOpen
  \bibfield  {author} {\bibinfo {author} {\bibfnamefont {Y.}~\bibnamefont
  {Wang}}, \bibinfo {author} {\bibfnamefont {L.}~\bibnamefont {Li}}, \bibinfo
  {author} {\bibfnamefont {W.}~\bibnamefont {Yao}}, \bibinfo {author}
  {\bibfnamefont {S.}~\bibnamefont {Song}}, \bibinfo {author} {\bibfnamefont
  {J.~T.}\ \bibnamefont {Sun}}, \bibinfo {author} {\bibfnamefont
  {J.}~\bibnamefont {Pan}}, \bibinfo {author} {\bibfnamefont {X.}~\bibnamefont
  {Ren}}, \bibinfo {author} {\bibfnamefont {C.}~\bibnamefont {Li}}, \bibinfo
  {author} {\bibfnamefont {E.}~\bibnamefont {Okunishi}}, \bibinfo {author}
  {\bibfnamefont {Y.-Q.}\ \bibnamefont {Wang}}, \bibinfo {author}
  {\bibfnamefont {E.}~\bibnamefont {Wang}}, \bibinfo {author} {\bibfnamefont
  {Y.}~\bibnamefont {Shao}}, \bibinfo {author} {\bibfnamefont {Y.~Y.}\
  \bibnamefont {Zhang}}, \bibinfo {author} {\bibfnamefont {H.-t.}\ \bibnamefont
  {Yang}}, \bibinfo {author} {\bibfnamefont {E.~F.}\ \bibnamefont {Schwier}},
  \bibinfo {author} {\bibfnamefont {H.}~\bibnamefont {Iwasawa}}, \bibinfo
  {author} {\bibfnamefont {K.}~\bibnamefont {Shimada}}, \bibinfo {author}
  {\bibfnamefont {M.}~\bibnamefont {Taniguchi}}, \bibinfo {author}
  {\bibfnamefont {Z.}~\bibnamefont {Cheng}}, \bibinfo {author} {\bibfnamefont
  {S.}~\bibnamefont {Zhou}}, \bibinfo {author} {\bibfnamefont {S.}~\bibnamefont
  {Du}}, \bibinfo {author} {\bibfnamefont {S.~J.}\ \bibnamefont {Pennycook}},
  \bibinfo {author} {\bibfnamefont {S.~T.}\ \bibnamefont {Pantelides}}, \ and\
  \bibinfo {author} {\bibfnamefont {H.-J.}\ \bibnamefont {Gao}},\ }\bibfield
  {title} {\enquote {\bibinfo {title} {Monolayer {PtSe}$_{\textrm{2}}$, a new
  semiconducting transition-metal-dichalcogenide, epitaxially grown by direct
  selenization of {Pt}},}\ }\href@noop {} {\bibfield  {journal} {\bibinfo
  {journal} {Nano Letters}\ }\textbf {\bibinfo {volume} {15}},\ \bibinfo
  {pages} {4013--4018} (\bibinfo {year} {2015})}\BibitemShut {NoStop}%
\bibitem [{\citenamefont {Zhao}\ \emph {et~al.}(2017)\citenamefont {Zhao},
  \citenamefont {Qiao}, \citenamefont {Yu}, \citenamefont {Yu}, \citenamefont
  {Xu}, \citenamefont {Lau}, \citenamefont {Zhou}, \citenamefont {Liu},
  \citenamefont {Wang}, \citenamefont {Ji} \emph {et~al.}}]{Zhao:2017}%
  \BibitemOpen
  \bibfield  {author} {\bibinfo {author} {\bibfnamefont {Y.}~\bibnamefont
  {Zhao}}, \bibinfo {author} {\bibfnamefont {J.}~\bibnamefont {Qiao}}, \bibinfo
  {author} {\bibfnamefont {Z.}~\bibnamefont {Yu}}, \bibinfo {author}
  {\bibfnamefont {P.}~\bibnamefont {Yu}}, \bibinfo {author} {\bibfnamefont
  {K.}~\bibnamefont {Xu}}, \bibinfo {author} {\bibfnamefont {S.~P.}\
  \bibnamefont {Lau}}, \bibinfo {author} {\bibfnamefont {W.}~\bibnamefont
  {Zhou}}, \bibinfo {author} {\bibfnamefont {Z.}~\bibnamefont {Liu}}, \bibinfo
  {author} {\bibfnamefont {X.}~\bibnamefont {Wang}}, \bibinfo {author}
  {\bibfnamefont {W.}~\bibnamefont {Ji}},  \emph {et~al.},\ }\bibfield  {title}
  {\enquote {\bibinfo {title} {High-electron-mobility and air-stable {2D}
  layered {PtSe}$_2$ {FET}s},}\ }\href@noop {} {\bibfield  {journal} {\bibinfo
  {journal} {Advanced Materials}\ }\textbf {\bibinfo {volume} {29}},\ \bibinfo
  {pages} {1604230} (\bibinfo {year} {2017})}\BibitemShut {NoStop}%
\bibitem [{\citenamefont {Yao}\ \emph {et~al.}(2017)\citenamefont {Yao},
  \citenamefont {Wang}, \citenamefont {Huang}, \citenamefont {Deng},
  \citenamefont {Yan}, \citenamefont {Zhang}, \citenamefont {Miyamoto},
  \citenamefont {Okuda}, \citenamefont {Li}, \citenamefont {Wang} \emph
  {et~al.}}]{Yao:2017}%
  \BibitemOpen
  \bibfield  {author} {\bibinfo {author} {\bibfnamefont {W.}~\bibnamefont
  {Yao}}, \bibinfo {author} {\bibfnamefont {E.}~\bibnamefont {Wang}}, \bibinfo
  {author} {\bibfnamefont {H.}~\bibnamefont {Huang}}, \bibinfo {author}
  {\bibfnamefont {K.}~\bibnamefont {Deng}}, \bibinfo {author} {\bibfnamefont
  {M.}~\bibnamefont {Yan}}, \bibinfo {author} {\bibfnamefont {K.}~\bibnamefont
  {Zhang}}, \bibinfo {author} {\bibfnamefont {K.}~\bibnamefont {Miyamoto}},
  \bibinfo {author} {\bibfnamefont {T.}~\bibnamefont {Okuda}}, \bibinfo
  {author} {\bibfnamefont {L.}~\bibnamefont {Li}}, \bibinfo {author}
  {\bibfnamefont {Y.}~\bibnamefont {Wang}},  \emph {et~al.},\ }\bibfield
  {title} {\enquote {\bibinfo {title} {Direct observation of spin-layer locking
  by local {Rashba} effect in monolayer semiconducting {PtSe}$_2$ film},}\
  }\href@noop {} {\bibfield  {journal} {\bibinfo  {journal} {Nature
  communications}\ }\textbf {\bibinfo {volume} {8}},\ \bibinfo {pages} {14216}
  (\bibinfo {year} {2017})}\BibitemShut {NoStop}%
\bibitem [{\citenamefont {Zhang}\ \emph {et~al.}(2014)\citenamefont {Zhang},
  \citenamefont {Liu}, \citenamefont {Luo}, \citenamefont {Freeman},\ and\
  \citenamefont {Zunger}}]{Zhang:2014}%
  \BibitemOpen
  \bibfield  {author} {\bibinfo {author} {\bibfnamefont {X.}~\bibnamefont
  {Zhang}}, \bibinfo {author} {\bibfnamefont {Q.}~\bibnamefont {Liu}}, \bibinfo
  {author} {\bibfnamefont {J.-W.}\ \bibnamefont {Luo}}, \bibinfo {author}
  {\bibfnamefont {A.~J.}\ \bibnamefont {Freeman}}, \ and\ \bibinfo {author}
  {\bibfnamefont {A.}~\bibnamefont {Zunger}},\ }\bibfield  {title} {\enquote
  {\bibinfo {title} {Hidden spin polarization in inversion-symmetric bulk
  crystals},}\ }\href@noop {} {\bibfield  {journal} {\bibinfo  {journal}
  {Nature Physics}\ }\textbf {\bibinfo {volume} {10}},\ \bibinfo {pages} {387}
  (\bibinfo {year} {2014})}\BibitemShut {NoStop}%
\bibitem [{\citenamefont {Yuan}\ \emph {et~al.}(2019)\citenamefont {Yuan},
  \citenamefont {Liu}, \citenamefont {Zhang}, \citenamefont {Luo},
  \citenamefont {Li},\ and\ \citenamefont {Zunger}}]{Yuan:2019}%
  \BibitemOpen
  \bibfield  {author} {\bibinfo {author} {\bibfnamefont {L.}~\bibnamefont
  {Yuan}}, \bibinfo {author} {\bibfnamefont {Q.}~\bibnamefont {Liu}}, \bibinfo
  {author} {\bibfnamefont {X.}~\bibnamefont {Zhang}}, \bibinfo {author}
  {\bibfnamefont {J.-W.}\ \bibnamefont {Luo}}, \bibinfo {author} {\bibfnamefont
  {S.-S.}\ \bibnamefont {Li}}, \ and\ \bibinfo {author} {\bibfnamefont
  {A.}~\bibnamefont {Zunger}},\ }\bibfield  {title} {\enquote {\bibinfo {title}
  {Uncovering and tailoring hidden {Rashba} spin--orbit splitting in
  centrosymmetric crystals},}\ }\href@noop {} {\bibfield  {journal} {\bibinfo
  {journal} {Nature communications}\ }\textbf {\bibinfo {volume} {10}},\
  \bibinfo {pages} {906} (\bibinfo {year} {2019})}\BibitemShut {NoStop}%
\bibitem [{\citenamefont {Absor}\ \emph {et~al.}(2018)\citenamefont {Absor},
  \citenamefont {Santoso}, \citenamefont {Harsojo}, \citenamefont {Abraha},
  \citenamefont {Kotaka}, \citenamefont {Ishii},\ and\ \citenamefont
  {Saito}}]{PtSe2_doping:2018}%
  \BibitemOpen
  \bibfield  {author} {\bibinfo {author} {\bibfnamefont {M.~A.~U.}\
  \bibnamefont {Absor}}, \bibinfo {author} {\bibfnamefont {I.}~\bibnamefont
  {Santoso}}, \bibinfo {author} {\bibnamefont {Harsojo}}, \bibinfo {author}
  {\bibfnamefont {K.}~\bibnamefont {Abraha}}, \bibinfo {author} {\bibfnamefont
  {H.}~\bibnamefont {Kotaka}}, \bibinfo {author} {\bibfnamefont
  {F.}~\bibnamefont {Ishii}}, \ and\ \bibinfo {author} {\bibfnamefont
  {M.}~\bibnamefont {Saito}},\ }\bibfield  {title} {\enquote {\bibinfo {title}
  {Strong {Rashba} effect in the localized impurity states of halogen-doped
  monolayer {PtSe}$_{2}$},}\ }\href@noop {} {\bibfield  {journal} {\bibinfo
  {journal} {Phys. Rev. B}\ }\textbf {\bibinfo {volume} {97}},\ \bibinfo
  {pages} {205138} (\bibinfo {year} {2018})}\BibitemShut {NoStop}%
\bibitem [{\citenamefont {Chang}\ \emph {et~al.}(2014)\citenamefont {Chang},
  \citenamefont {Zhang}, \citenamefont {Zhu}, \citenamefont {Han},
  \citenamefont {Pu}, \citenamefont {Chang}, \citenamefont {Hsu}, \citenamefont
  {Huang}, \citenamefont {Hsu}, \citenamefont {Chiu} \emph
  {et~al.}}]{Chang:2014}%
  \BibitemOpen
  \bibfield  {author} {\bibinfo {author} {\bibfnamefont {Y.-H.}\ \bibnamefont
  {Chang}}, \bibinfo {author} {\bibfnamefont {W.}~\bibnamefont {Zhang}},
  \bibinfo {author} {\bibfnamefont {Y.}~\bibnamefont {Zhu}}, \bibinfo {author}
  {\bibfnamefont {Y.}~\bibnamefont {Han}}, \bibinfo {author} {\bibfnamefont
  {J.}~\bibnamefont {Pu}}, \bibinfo {author} {\bibfnamefont {J.-K.}\
  \bibnamefont {Chang}}, \bibinfo {author} {\bibfnamefont {W.-T.}\ \bibnamefont
  {Hsu}}, \bibinfo {author} {\bibfnamefont {J.-K.}\ \bibnamefont {Huang}},
  \bibinfo {author} {\bibfnamefont {C.-L.}\ \bibnamefont {Hsu}}, \bibinfo
  {author} {\bibfnamefont {M.-H.}\ \bibnamefont {Chiu}},  \emph {et~al.},\
  }\bibfield  {title} {\enquote {\bibinfo {title} {Monolayer {MoSe}$_2$ grown
  by chemical vapor deposition for fast photodetection},}\ }\href@noop {}
  {\bibfield  {journal} {\bibinfo  {journal} {ACS nano}\ }\textbf {\bibinfo
  {volume} {8}},\ \bibinfo {pages} {8582--8590} (\bibinfo {year}
  {2014})}\BibitemShut {NoStop}%
\bibitem [{\citenamefont {Reyes-Retana}\ and\ \citenamefont
  {Cervantes-Sodi}(2016)}]{Reyes:2016}%
  \BibitemOpen
  \bibfield  {author} {\bibinfo {author} {\bibfnamefont {J.}~\bibnamefont
  {Reyes-Retana}}\ and\ \bibinfo {author} {\bibfnamefont {F.}~\bibnamefont
  {Cervantes-Sodi}},\ }\bibfield  {title} {\enquote {\bibinfo {title}
  {Spin-orbital effects in metal-dichalcogenide semiconducting monolayers},}\
  }\href@noop {} {\bibfield  {journal} {\bibinfo  {journal} {Scientific
  reports}\ }\textbf {\bibinfo {volume} {6}},\ \bibinfo {pages} {24093}
  (\bibinfo {year} {2016})}\BibitemShut {NoStop}%
\bibitem [{\citenamefont {Kresse}\ and\ \citenamefont
  {Joubert}(1999)}]{Kresse:1999}%
  \BibitemOpen
  \bibfield  {author} {\bibinfo {author} {\bibfnamefont {G.}~\bibnamefont
  {Kresse}}\ and\ \bibinfo {author} {\bibfnamefont {D.}~\bibnamefont
  {Joubert}},\ }\bibfield  {title} {\enquote {\bibinfo {title} {From ultrasoft
  pseudopotentials to the projector augmented-wave method},}\ }\href@noop {}
  {\bibfield  {journal} {\bibinfo  {journal} {Phys. Rev. B}\ }\textbf {\bibinfo
  {volume} {59}},\ \bibinfo {pages} {1758--1775} (\bibinfo {year}
  {1999})}\BibitemShut {NoStop}%
\bibitem [{\citenamefont {Perdew}, \citenamefont {Burke},\ and\ \citenamefont
  {Ernzerhof}(1996)}]{Perdew:1996}%
  \BibitemOpen
  \bibfield  {author} {\bibinfo {author} {\bibfnamefont {J.~P.}\ \bibnamefont
  {Perdew}}, \bibinfo {author} {\bibfnamefont {K.}~\bibnamefont {Burke}}, \
  and\ \bibinfo {author} {\bibfnamefont {M.}~\bibnamefont {Ernzerhof}},\
  }\bibfield  {title} {\enquote {\bibinfo {title} {Generalized gradient
  approximation made simple},}\ }\href@noop {} {\bibfield  {journal} {\bibinfo
  {journal} {Phys. Rev. Lett.}\ }\textbf {\bibinfo {volume} {77}},\ \bibinfo
  {pages} {3865--3868} (\bibinfo {year} {1996})}\BibitemShut {NoStop}%
\bibitem [{\citenamefont {Grimme}\ \emph {et~al.}(2010)\citenamefont {Grimme},
  \citenamefont {Antony}, \citenamefont {Ehrlich},\ and\ \citenamefont
  {Krieg}}]{Grimme:2010}%
  \BibitemOpen
  \bibfield  {author} {\bibinfo {author} {\bibfnamefont {S.}~\bibnamefont
  {Grimme}}, \bibinfo {author} {\bibfnamefont {J.}~\bibnamefont {Antony}},
  \bibinfo {author} {\bibfnamefont {S.}~\bibnamefont {Ehrlich}}, \ and\
  \bibinfo {author} {\bibfnamefont {H.}~\bibnamefont {Krieg}},\ }\bibfield
  {title} {\enquote {\bibinfo {title} {A consistent and accurate \textit{ab
  initio} parametrization of density functional dispersion correction ({DFT-D})
  for the 94 elements {H}-{Pu}},}\ }\href@noop {} {\bibfield  {journal}
  {\bibinfo  {journal} {The Journal of Chemical Physics}\ }\textbf {\bibinfo
  {volume} {132}},\ \bibinfo {pages} {154104} (\bibinfo {year}
  {2010})}\BibitemShut {NoStop}%
\bibitem [{\citenamefont {Zhu}, \citenamefont {Cheng},\ and\ \citenamefont
  {Schwingenschl\"ogl}(2011)}]{Schwingen:2011}%
  \BibitemOpen
  \bibfield  {author} {\bibinfo {author} {\bibfnamefont {Z.~Y.}\ \bibnamefont
  {Zhu}}, \bibinfo {author} {\bibfnamefont {Y.~C.}\ \bibnamefont {Cheng}}, \
  and\ \bibinfo {author} {\bibfnamefont {U.}~\bibnamefont
  {Schwingenschl\"ogl}},\ }\bibfield  {title} {\enquote {\bibinfo {title}
  {Giant spin-orbit-induced spin splitting in two-dimensional transition-metal
  dichalcogenide semiconductors},}\ }\href@noop {} {\bibfield  {journal}
  {\bibinfo  {journal} {Phys. Rev. B}\ }\textbf {\bibinfo {volume} {84}},\
  \bibinfo {pages} {153402} (\bibinfo {year} {2011})}\BibitemShut {NoStop}%
\bibitem [{\citenamefont {Kang}\ \emph {et~al.}(2013)\citenamefont {Kang},
  \citenamefont {Li}, \citenamefont {Li}, \citenamefont {Xia},\ and\
  \citenamefont {Wang}}]{Kang:2013}%
  \BibitemOpen
  \bibfield  {author} {\bibinfo {author} {\bibfnamefont {J.}~\bibnamefont
  {Kang}}, \bibinfo {author} {\bibfnamefont {J.}~\bibnamefont {Li}}, \bibinfo
  {author} {\bibfnamefont {S.-S.}\ \bibnamefont {Li}}, \bibinfo {author}
  {\bibfnamefont {J.-B.}\ \bibnamefont {Xia}}, \ and\ \bibinfo {author}
  {\bibfnamefont {L.-W.}\ \bibnamefont {Wang}},\ }\bibfield  {title} {\enquote
  {\bibinfo {title} {Electronic structural moir\'{e} pattern effects on
  {MoS}$_2$/{MoSe}$_2$ 2{D} heterostructures},}\ }\href@noop {} {\bibfield
  {journal} {\bibinfo  {journal} {Nano Lett.}\ }\textbf {\bibinfo {volume}
  {13}} (\bibinfo {year} {2013})}\BibitemShut {NoStop}%
\bibitem [{\citenamefont {Komsa}\ and\ \citenamefont
  {Krasheninnikov}(2013)}]{Komsa:2013}%
  \BibitemOpen
  \bibfield  {author} {\bibinfo {author} {\bibfnamefont {H.-P.}\ \bibnamefont
  {Komsa}}\ and\ \bibinfo {author} {\bibfnamefont {A.~V.}\ \bibnamefont
  {Krasheninnikov}},\ }\bibfield  {title} {\enquote {\bibinfo {title}
  {Electronic structures and optical properties of realistic transition metal
  dichalcogenide heterostructures from first principles},}\ }\href@noop {}
  {\bibfield  {journal} {\bibinfo  {journal} {Phys. Rev. B}\ }\textbf {\bibinfo
  {volume} {88}},\ \bibinfo {pages} {085318} (\bibinfo {year}
  {2013})}\BibitemShut {NoStop}%
\bibitem [{\citenamefont {Su}\ \emph {et~al.}(2016)\citenamefont {Su},
  \citenamefont {Ju}, \citenamefont {Zhang}, \citenamefont {Guo}, \citenamefont
  {Zheng}, \citenamefont {Yong},\ and\ \citenamefont {Li}}]{Su:2016}%
  \BibitemOpen
  \bibfield  {author} {\bibinfo {author} {\bibfnamefont {X.}~\bibnamefont
  {Su}}, \bibinfo {author} {\bibfnamefont {W.}~\bibnamefont {Ju}}, \bibinfo
  {author} {\bibfnamefont {R.}~\bibnamefont {Zhang}}, \bibinfo {author}
  {\bibfnamefont {C.}~\bibnamefont {Guo}}, \bibinfo {author} {\bibfnamefont
  {J.}~\bibnamefont {Zheng}}, \bibinfo {author} {\bibfnamefont
  {Y.}~\bibnamefont {Yong}}, \ and\ \bibinfo {author} {\bibfnamefont
  {X.}~\bibnamefont {Li}},\ }\bibfield  {title} {\enquote {\bibinfo {title}
  {Bandgap engineering of {MoS}$_2$/{MX}$_2$ ({MX}$_2$={WS}$_2$, {MoSe}$_2$ and
  {WSe}$_2$) heterobilayers subjected to biaxial strain and normal compressive
  strain},}\ }\href@noop {} {\bibfield  {journal} {\bibinfo  {journal} {RSC
  Advances}\ }\textbf {\bibinfo {volume} {6}},\ \bibinfo {pages} {18319--18325}
  (\bibinfo {year} {2016})}\BibitemShut {NoStop}%
\bibitem [{\citenamefont {Rybkovskiy}, \citenamefont {Osadchy},\ and\
  \citenamefont {Obraztsova}(2014)}]{Rybkovskiy2014}%
  \BibitemOpen
  \bibfield  {author} {\bibinfo {author} {\bibfnamefont {D.~V.}\ \bibnamefont
  {Rybkovskiy}}, \bibinfo {author} {\bibfnamefont {A.~V.}\ \bibnamefont
  {Osadchy}}, \ and\ \bibinfo {author} {\bibfnamefont {E.~D.}\ \bibnamefont
  {Obraztsova}},\ }\bibfield  {title} {\enquote {\bibinfo {title} {Transition
  from parabolic to ring-shaped valence band maximum in few-layer gas, gase,
  and inse},}\ }\href {\doibase 10.1103/PhysRevB.90.235302} {\bibfield
  {journal} {\bibinfo  {journal} {Phys. Rev. B}\ }\textbf {\bibinfo {volume}
  {90}},\ \bibinfo {pages} {235302} (\bibinfo {year} {2014})}\BibitemShut
  {NoStop}%
\bibitem [{\citenamefont {Chuang}\ \emph {et~al.}(2015)\citenamefont {Chuang},
  \citenamefont {Ho}, \citenamefont {Smith}, \citenamefont {Sfigakis},
  \citenamefont {Pepper}, \citenamefont {Chen}, \citenamefont {Fan},
  \citenamefont {Griffiths}, \citenamefont {Farrer}, \citenamefont {Beere},
  \citenamefont {Jones}, \citenamefont {Ritchie},\ and\ \citenamefont
  {Chen}}]{All_electric:2015}%
  \BibitemOpen
  \bibfield  {author} {\bibinfo {author} {\bibfnamefont {P.}~\bibnamefont
  {Chuang}}, \bibinfo {author} {\bibfnamefont {S.-C.}\ \bibnamefont {Ho}},
  \bibinfo {author} {\bibfnamefont {L.~W.}\ \bibnamefont {Smith}}, \bibinfo
  {author} {\bibfnamefont {F.}~\bibnamefont {Sfigakis}}, \bibinfo {author}
  {\bibfnamefont {M.}~\bibnamefont {Pepper}}, \bibinfo {author} {\bibfnamefont
  {C.-H.}\ \bibnamefont {Chen}}, \bibinfo {author} {\bibfnamefont {J.-C.}\
  \bibnamefont {Fan}}, \bibinfo {author} {\bibfnamefont {J.~P.}\ \bibnamefont
  {Griffiths}}, \bibinfo {author} {\bibfnamefont {I.}~\bibnamefont {Farrer}},
  \bibinfo {author} {\bibfnamefont {H.~E.}\ \bibnamefont {Beere}}, \bibinfo
  {author} {\bibfnamefont {G.~A.~C.}\ \bibnamefont {Jones}}, \bibinfo {author}
  {\bibfnamefont {D.~A.}\ \bibnamefont {Ritchie}}, \ and\ \bibinfo {author}
  {\bibfnamefont {T.-M.}\ \bibnamefont {Chen}},\ }\bibfield  {title} {\enquote
  {\bibinfo {title} {All-electric all-semiconductor spin field-effect
  transistors},}\ }\href@noop {} {\bibfield  {journal} {\bibinfo  {journal}
  {Nature Nanotechnology}\ }\textbf {\bibinfo {volume} {10}},\ \bibinfo {pages}
  {35--39} (\bibinfo {year} {2015})}\BibitemShut {NoStop}%
\bibitem [{\citenamefont {Luo}\ \emph {et~al.}(2017)\citenamefont {Luo},
  \citenamefont {Xu}, \citenamefont {Zhu}, \citenamefont {Wu}, \citenamefont
  {McCormick}, \citenamefont {Zhan}, \citenamefont {Neupane},\ and\
  \citenamefont {Kawakami}}]{Luo:2017}%
  \BibitemOpen
  \bibfield  {author} {\bibinfo {author} {\bibfnamefont {Y.~K.}\ \bibnamefont
  {Luo}}, \bibinfo {author} {\bibfnamefont {J.}~\bibnamefont {Xu}}, \bibinfo
  {author} {\bibfnamefont {T.}~\bibnamefont {Zhu}}, \bibinfo {author}
  {\bibfnamefont {G.}~\bibnamefont {Wu}}, \bibinfo {author} {\bibfnamefont
  {E.~J.}\ \bibnamefont {McCormick}}, \bibinfo {author} {\bibfnamefont
  {W.}~\bibnamefont {Zhan}}, \bibinfo {author} {\bibfnamefont {M.~R.}\
  \bibnamefont {Neupane}}, \ and\ \bibinfo {author} {\bibfnamefont {R.~K.}\
  \bibnamefont {Kawakami}},\ }\bibfield  {title} {\enquote {\bibinfo {title}
  {Opto-valleytronic spin injection in monolayer {MoS}$_2$/few-layer graphene
  hybrid spin valves},}\ }\href@noop {} {\bibfield  {journal} {\bibinfo
  {journal} {Nano letters}\ }\textbf {\bibinfo {volume} {17}},\ \bibinfo
  {pages} {3877--3883} (\bibinfo {year} {2017})}\BibitemShut {NoStop}%
\bibitem [{Note1()}]{Note1}%
  \BibitemOpen
  \bibinfo {note} {The change of the effective mass $m$ is minor, we use the
  average value $m=0.81 m_e$ in our estimation.}\BibitemShut {Stop}%
\end{thebibliography}%

\end{document}



\title{Supplementary Materials for "Tunable Giant Rashba-type Spin Splitting in PtSe$_2$/MoSe$_2$ Heterostructure"}
\author{Longjun Xiang}
\author{Youqi Ke}
\email{keyq@shanghaitech.edu.cn}
\author{Qingyun Zhang}
\email{zhangqy2@shanghaitech.edu.cn}
\affiliation{School of Physical Science and Technology, ShanghaiTech University, Shanghai, 201210, China}
\affiliation{University of Chinese Academy of Sciences, Beijing 100049, China}
\date{\today}

\maketitle 

\section{Structure information}

\begin{center}
\begin{table}[htb!]
\begin{tabular}{l|c|c|c}
\hline
        & $\delta (\%)$ & $a_0 (\mathrm{\AA})$ & $d_i (\mathrm{\AA})$  \\
\hline
PtS$_2$/MoS$_2$  & $2.8$ & $6.268$              & $3.206$             \\
PtSe$_2$/MoSe$_2$& $0.4$ & $6.534$              & $3.228$             \\
PtTe$_2$/MoTe$_2$& $2.0$ & $6.979$              & $3.505$             \\
PtS$_2$/WS$_2$   & $2.7$ & $6.274$              & $3.203$             \\
PtSe$_2$/WSe$_2$ & $1.7$ & $6.531$              & $3.246$             \\
PtTe$_2$/WTe$_2$ & $2.2$ & $6.981$              & $3.514$             \\
\hline
\end{tabular}
\caption{ \label{table} Lattice mismatch, relaxed lattice constant $a_0$ and interlayer distances $d_i$
for all six heterostructures PtX$_2$/MX$_2$ (M=Mo, W; X=S, Se, Te).}
\end{table}
\end{center}

All six heterostructures PtX$_2$/MX$_2$ (M=Mo, W; X=S, Se, Te) are built with
$\sqrt{3}\times\sqrt{3}\times1$ R$30^{\circ}$ PtX$_2$ monolayer and $2\times2\times1$
MX$_2$ monolayer, respectively. For completeness, we list the information about
lattice mismatch, relaxed lattice constant and layer distance in TABLE \ref{table}.

\section{Generalized Rashba Hamiltonian}

\begin{figure}[htb!]
\centering \includegraphics[width=0.98\columnwidth]{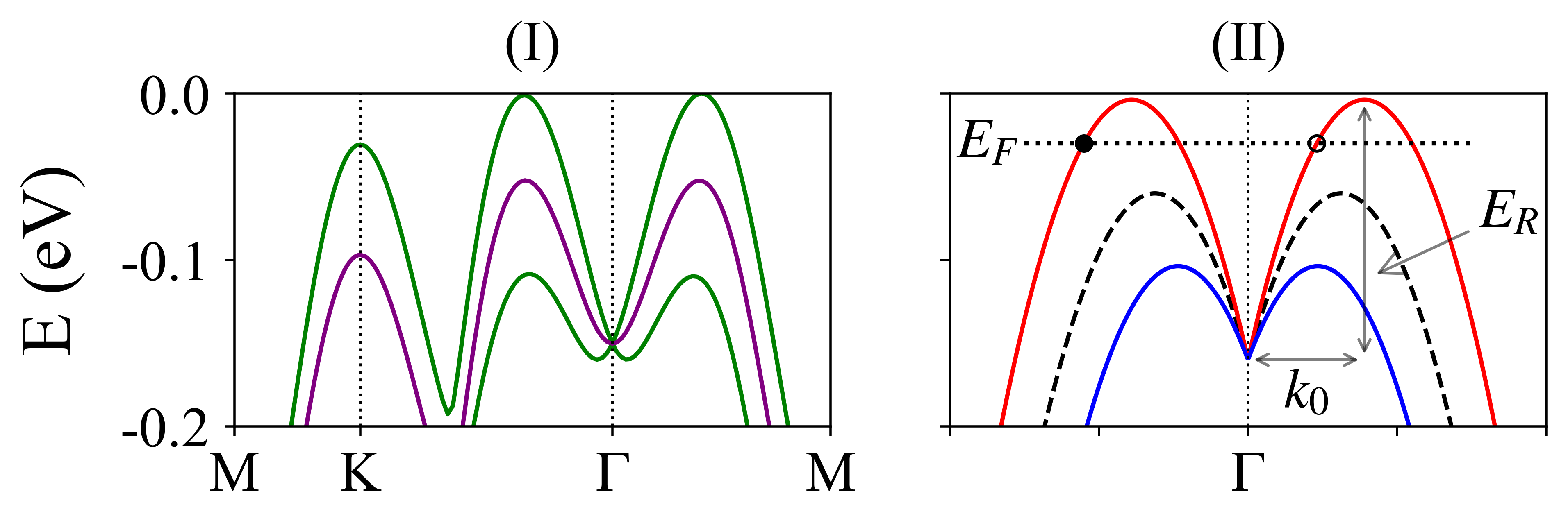}
\caption{\label{hamiltonian} (I) Band structures of the heterostructure PtSe$_2$/MoSe$_2$
without (purple) and with (green) SOC. (II) Band structures without SOC (dashed line)
and with SOC obtained from the generalized Hamiltonian $H_0$ and $H$ with $k_y=0$, respectively.
} 
\end{figure}

We notice that the valence band maximum (VBM) around $\Gamma$ point show the
so-called “sombrero hat” dispersion\cite{Rybkovskiy2014} 
before switching on the SOC (see the purple line in FIG.\ref{hamiltonian} (I)),
which is different from the typical parabolic dispersion of two-dimensional 
electron gas. This feature is caused by the spin-independent interaction with crystal field.
To well describe this feature, we can introduce 
the spin-independent Hamiltonian:

\begin{equation}
H_0=-\frac{\hbar^2(k_x^2+k_y^2)}{2m}+c\sqrt{k_x^2+k_y^2}.
\end{equation}
The band for $H_0$ with $k_y=0$ is shown below by the black dashed line in FIG.\ref{hamiltonian} (II),
which shows a good approximation for the VBM around $\Gamma$ point without SOC in PtSe$_2$/MoSe$_2$.
Taking the Rashba SOC into account, then the Hamiltonian reads:

\begin{equation}
H=-\frac{\hbar^2\left(k_x^2+k_y^2\right)}{2m}+c\sqrt{k_x^2+k_y^2}+
\alpha_R \left(\sigma_x \cdot k_y-\sigma_y \cdot k_x\right)
\end{equation}

The dashed band in FIG.\ref{hamiltonian} (II) is split to the upper (red) and the lower (blue) band,
in which the red upper band also shows a good approximation for the upper spit band in PtSe$_2$/MoSe$_2$. 
This Hamiltonian is referred as generalized Rashba Hamiltonian. In analogy with Datta and Das’s original
proposal, we provide some further derivation to show how the upper split band described by the generalized
Rashba Hamiltonian can also be utilized in SFET. From the generalized Rashba Hamiltonian $H$ it is easy to
obtain two sub-bands by solving the corresponding eigenvalue problem:
\begin{align}
E_1 & = -\frac{\hbar^2(k_x^2+k_y^2)}{2m}+(c+\alpha_R)\sqrt{k_x^2+k_y^2} \\
E_2 & = -\frac{\hbar^2(k_x^2+k_y^2)}{2m}+(c-\alpha_R)\sqrt{k_x^2+k_y^2}
\end{align}
Here $E_1$ and $E_2$ represents the upper and lower sub-bands in the two-dimensional $k_x$-$k_y$ plane, respectively.
Correspondingly, the eigenvectors are
\begin{align}
|\psi_1\rangle =
\begin{bmatrix}
ie^{-i\theta_k} \\
1
\end{bmatrix}
\quad \text{and} \quad
|\psi_2\rangle =
\begin{bmatrix}
-ie^{-i\theta_k} \\
1
\end{bmatrix}
\end{align}
in which $\theta_k={tg}^{-1}\frac{k_y}{k_x}$ reflects the spin-momentum locking property.
In the following, we focus on the upper sub-band since it shows similarity to that given by
 original Rashba Hamiltonian. In the following, we define a generalized Rashba constant $\eta_R=c+\alpha_R$. 
Without loss of generality, we will consider the transport along $x$-axis ($k_y=0$).
When the Fermi level crosses only the upper sub-band, the two modes with $+y$ and $–y$ polarized spins,
traveling along $+x$ direction, are marked by the respective black dot and circle in FIG.\ref{hamiltonian} (II),
and the corresponding wave-vectors are obtained as
\begin{align}
k_{x1} &=+\frac{m\eta_R}{\hbar^2}-\sqrt{\frac{m^2{\eta_R}^2}{\hbar^4}-\frac{2m}{\hbar^2}E_F} \\
k_{x2} &=-\frac{m\eta_R}{\hbar^2}-\sqrt{\frac{m^2{\eta_R}^2}{\hbar^4}-\frac{2m}{\hbar^2}E_F}
\end{align}
Then a differential phase shift is introduced between two modes when traveling through the device:
\begin{equation}
\Delta{\theta} = (k_{x1}-k_{x2})L=\frac{2m\eta_{R}L}{\hbar^2}
\end{equation}
where $L$ is the distance between spin injector and detector. If we suppose the injected spin is
along $+z$ direction, which can be represented as a linear combination of the above two transport
modes with $+y$ and $-y$ spin polarizations:
\begin{equation}
\begin{bmatrix}
1\\
0
\end{bmatrix}
=
\frac{1}{2i}|\psi_1^{+}\rangle -\frac{1}{2i} |\psi_1^{-}\rangle
=
\frac{1}{2i}
\begin{bmatrix}
i\\
1
\end{bmatrix}
-
\frac{1}{2i}
\begin{bmatrix}
-i\\
1
\end{bmatrix}
\end{equation}
When the two spins arrived at the detector, the mode with $|\psi_1^{+}\rangle$ goes through a
phase $\theta_1=k_{x1}L$, while the mode with $|\psi_1^{-}\rangle$ acquires a phase $\theta_2=k_{x2}L$.
Now the final state is:
\begin{equation}
\frac{1}{2i}
\begin{bmatrix}
i\\
1
\end{bmatrix}
e^{i\theta_1}
-
\frac{1}{2i}
\begin{bmatrix}
-i\\
1
\end{bmatrix}
e^{i\theta_2}
=
\frac{1}{2i}
\begin{bmatrix}
ie^{i\Delta{\theta}}+i\\
e^{i\Delta{\theta}}-1
\end{bmatrix}
e^{i\theta_2}
\end{equation}
Here the differential phase shift is $\Delta{\theta}=\theta_1-\theta_2=2m\eta_{R}L/\hbar^2$.
If we take $\Delta{\theta}=\pi$, the final state becomes
\begin{equation*}
\begin{bmatrix}
0\\
1
\end{bmatrix}
e^{i\theta_2}
\end{equation*}
with $-z$ spin polarization, which means the spin is flipped. It is worth noticing that
the differential phase shift is proportional to $\eta_R$, which is caused by both SOC as well as crystal field. 
It is different from the original idea of Datta and Das based on two-dimensional electron gas.

For PtSe$_2$/MoSe$_2$, the generalized Rashba constant $\eta_R$ can be tuned effectively by electric field and strain.
Therefore, the energy interval from 0 eV to -0.1 eV around $\Gamma$ point in PtSe$_2$/MoSe$_2$ is
suitable for SFET application. In particular,  $\eta_R$ can still be estimated by  $2E_R/k_0$ with $E_R$
the generalized Rashba energy and $k_0$ the momentum offset, as shown in FIG.\ref{hamiltonian} (II).

If the Fermi level goes below $-$0.1 eV in PtSe$_2$/MoSe$_2$ then two problems will appear:
(1) states at the valence band edge around $K$ points will fall into this energy range; (2) the strong
energy dependence of $\Delta{\theta}$. Therefore, the states below $-$0.1 eV are not utilized
in the proposed SFET based on PtSe$_2$/MoSe$_2$.

\section{Orbital hybridization}

\begin{figure}[htb!]
\centering \includegraphics[width=0.98\columnwidth]{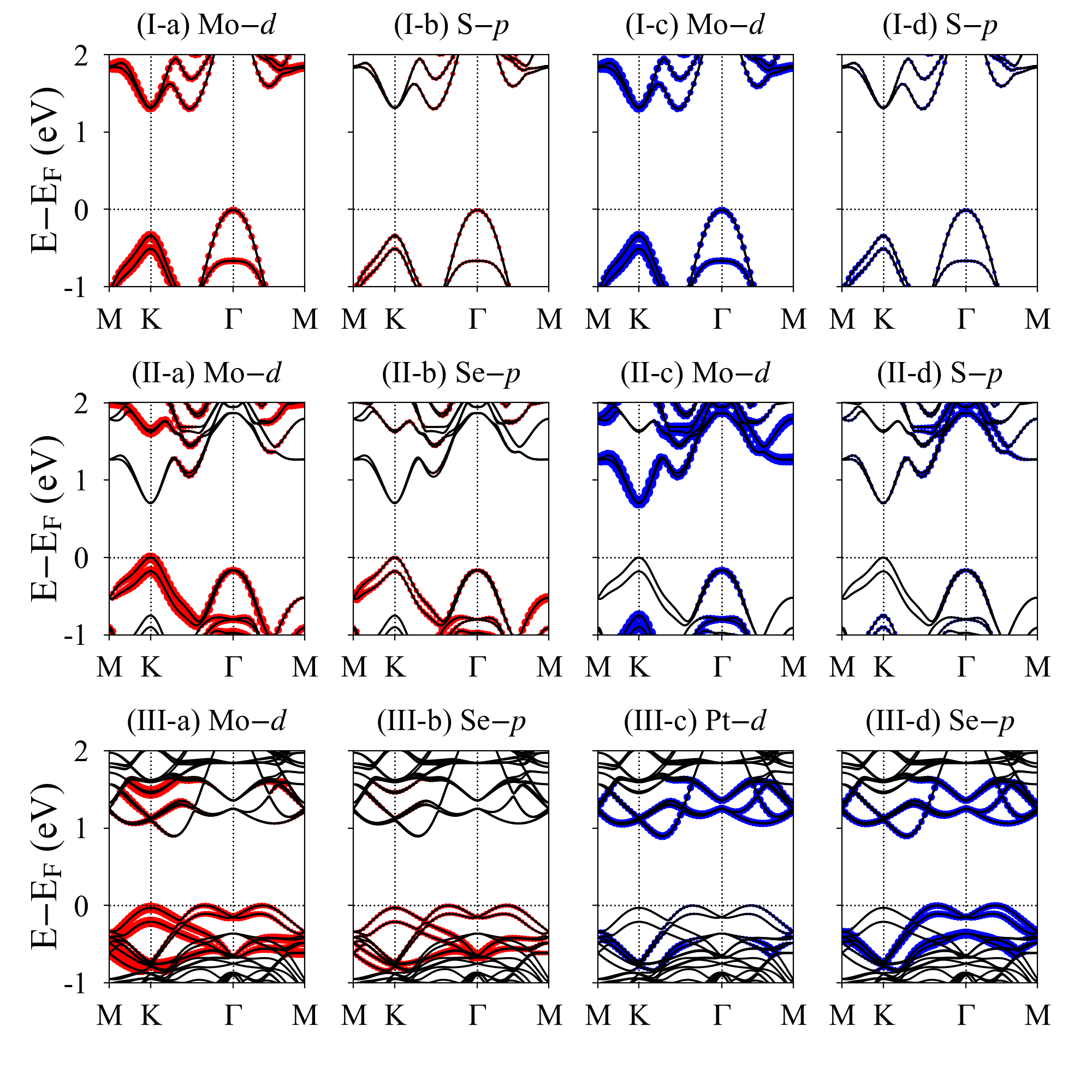}
\caption{\label{orbital1}
Band structures of (I) bilayer MoS$_2$, (II) MoSe$_2$/MoS$_2$ and (III) PtSe$_2$/MoSe$_2$
heterostructures with projections to the corresponding atomic orbitals for the bands near Fermi level.
Here the size of circles represents the contribution of atomic orbitals. The red and blue colors
represent two constituent layers. In the following, we adopt the same convention.}
\end{figure}

\begin{figure}[htb!]
\centering \includegraphics[width=0.98\columnwidth]{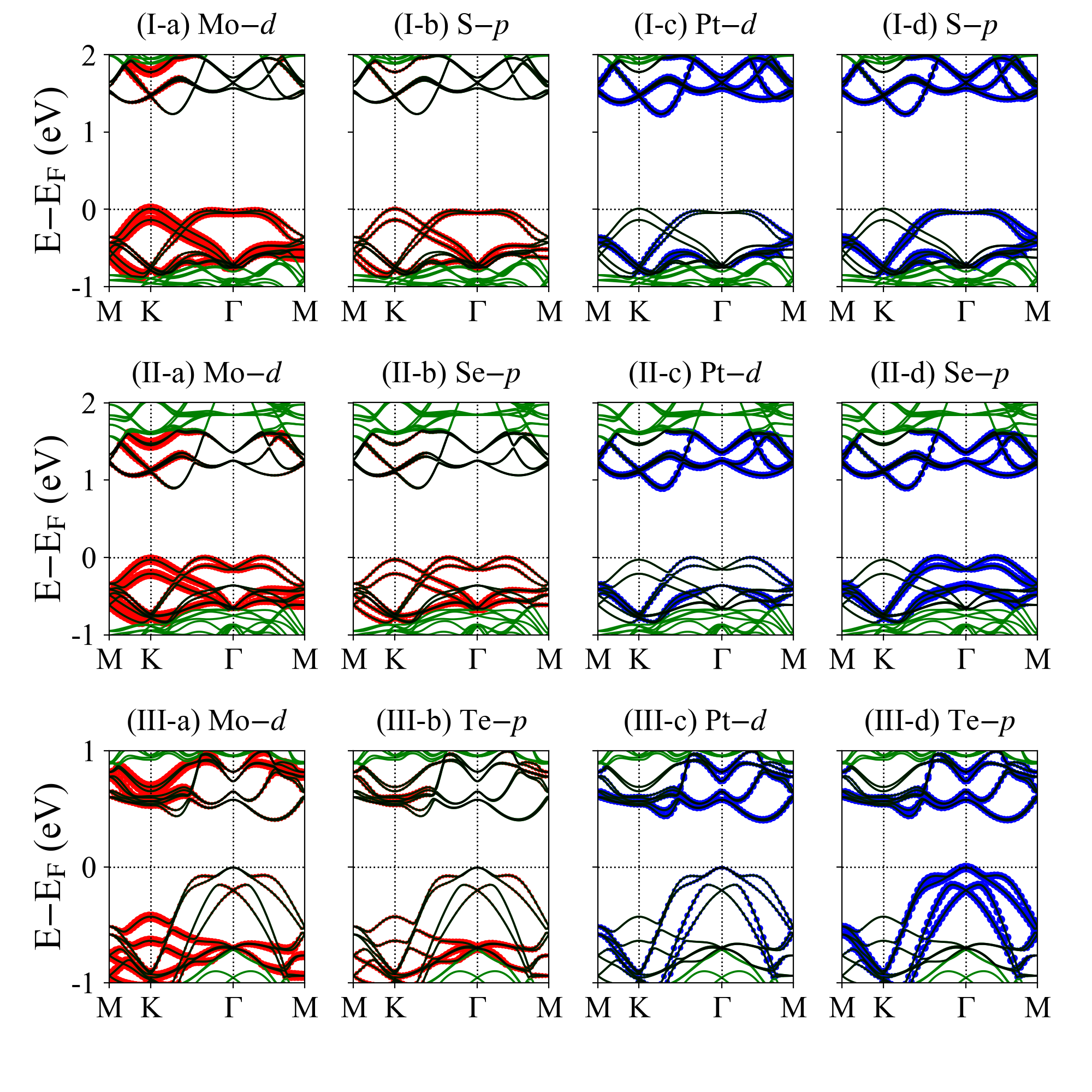}
\caption{\label{orbital2}
The band structures of (I) PtS$_2$/MoS$_2$, (II) PtSe$_2$/MoSe$_2$ and (III) PtTe$_2$/MoTe$_2$
heterostructures with projections to the corresponding atomic orbitals for the bands near Fermi level.}
\end{figure}

\begin{figure}[htb!]
\centering \includegraphics[width=0.98\columnwidth]{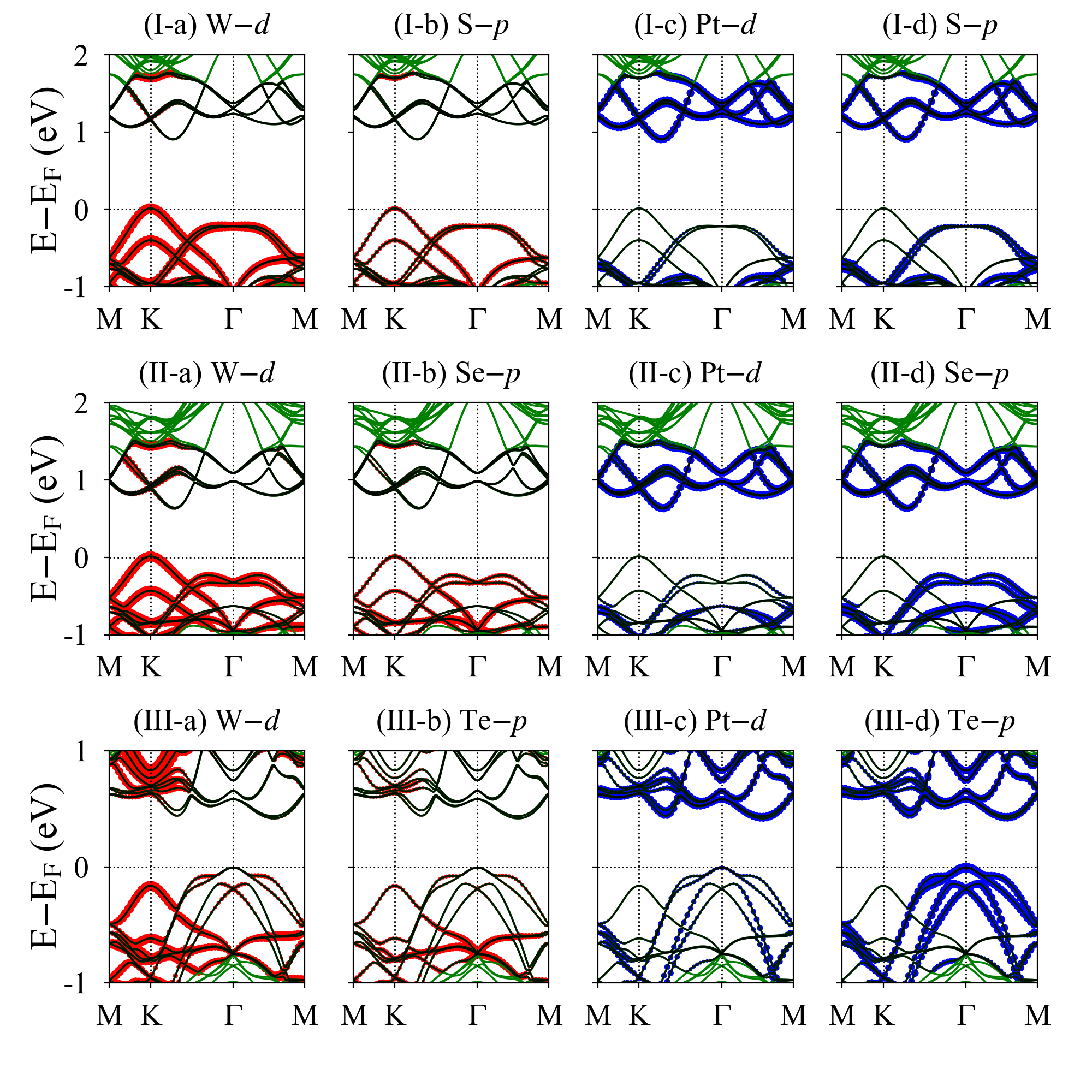}
\caption{\label{orbital3}
Band structures of (I) PtS$_2$/WS$_2$, (II) PtSe$_2$/WSe$_2$ and (III) PtTe$_2$/WTe$_2$
heterostructures with projections to the corresponding atomic orbitals for the bands near Fermi level.}
\end{figure}

In FIG.\ref{orbital1}, we compare the features of orbital hybridization around $\Gamma$ point for the
atomic-orbitals in bilayer MoS$_2$, MoSe$_2$/MoS$_2$ and PtSe$_2$/MoSe$_2$. It is found
that the hybridization in bilayer MoS$_2$ and MoSe$_2$/MoS$_2$ is dominated by the Mo-$d$ orbitals
from different layers, as shown in FIG.\ref{orbital1} (I-a), (I-c), (II-a) and (II-c). However, the
hybridization around $\Gamma$ point in PtSe$_2$/MoSe$_2$ is mainly contributed by Mo-$d$ orbitals
in MoSe$_2$ layer and Se-$p$ orbitals in PtSe$_2$, which shows a distinct feature compared to bilayer
MoS$_2$ and MoS$_2$/MoSe$_2$. This emerging hybridization results in the giant Rashba-type spin
splitting in PtSe$_2$/MoSe$_2$, as discussed in the main manuscript.

In FIG.\ref{orbital2} and FIG.\ref{orbital3}, we present the band structures and features
of orbital hybridization for the PtX$_2$/MoX$_2$ and PtX$_2$/WX$_2$ (X=S, Se, Te), respectively.
As shown in FIG.\ref{orbital2} (I) and (II),
we observe similar hybridization features for PtS$_2$/MoS$_2$ and PtSe$_2$/MoSe$_2$, but the
band splitting around $\Gamma$ point in PtS$_2$/MoS$_2$ is much smaller due to the weaker
spin-orbit coupling in S atom. While for PtTe$_2$/MoTe$_2$ shown in FIG.\ref{orbital2} (III), almost no
$p$-$d$ hybridization and no band splitting around $\Gamma$ point is observed at VBE. Substituting
Mo atoms with W atoms in these heterostructures will bring about two main changes to their band
structures, as shown in FIG.\ref{orbital3}. Firstly, the band splitting of VBE at $K$ point become larger due
to the stronger SOC in W atom than that in Mo atom. Secondly, the energy levels of VBE around
$\Gamma$ point are suppressed compared to those at $K$ point. However, the orbital hybridizations
in PtX$_2$/WX$_2$ are almost the same as in PtX$_2$/MoX$_2$.

\section{Channel Length}

The two band structures with $\pm$ 1.5 eV/nm under 1.5\% tensile strain is utilized
to estimate the minimum channel length. The magnitude of average effective mass for
valence band edge around $\Gamma$ point is $m=0.81 m_e$. And the generalized Rashba
constant $\eta_R$ for $-$1.5 V/nm and +1.5 V/nm are 1.098 eV$\cdot \mathrm{\AA}$ and
0.890 eV$\cdot \mathrm{\AA}$, respectively. If we set $\Delta\theta=\pi$, then channel
length is calculated as: $$L=\frac{\Delta \theta \hbar^2}{2m\Delta\eta_R}=7.1 nm$$